\documentclass[12pt]{article}

\usepackage[usenames,dvipsnames]{xcolor}

\usepackage{tikz}
\usetikzlibrary{matrix,arrows,positioning,shapes,snakes,fadings,decorations.pathmorphing,
decorations.pathreplacing,automata,shadows,decorations.markings,patterns,decorations.text}
\usepackage{jheppub}
\usepackage{amsmath,amssymb,euscript,array,mathrsfs,appendix,ctable,mathabx}
\usepackage{mathtools,float}
\usepackage{arydshln}
\usepackage{todonotes}
\usepackage{graphicx}
\usepackage{wrapfig}
\usepackage{enumitem}
\usepackage{multirow}
\usepackage{bm,upgreek}
\usepackage{stmaryrd}

\usepackage{caption}
\usepackage{subcaption}

\usepackage{amsthm}

\newcommand*\circled[1]{{\footnotesize\tikz[baseline=(char.base)]{%
            \node[shape=circle,fill=black!20,draw,inner sep=1pt] (char) {#1};}}}

\newcommand \arXiv [1]{\href{http://arxiv.org/abs/#1}{\tt arXiv:#1}}

\usepackage{titlesec}
\titleformat{\subsection}[display]{\it}{}{0.1cm}{\vspace{-1.5cm}\begin{center}\thesubsection\hspace{0.2cm}}[\end{center}\vspace{-0.5cm}]

\newcommand{\ext}{\text{ext}}

\newcommand\cc{c}

\newcommand{\EQ}[1]{\begin{equation}\begin{split} #1
\end{split}\end{equation}}

\title{Black Hole Information Recovery in JT Gravity}
\author{Zsolt Gyongyosi, Timothy J. Hollowood, S.~Prem Kumar, Andrea Legramandi and Neil Talwar}
\affiliation{Department of Physics, Swansea University, Singleton Park, Swansea, U.K.}
\emailAdd{z.gyongyosi.2133547@swansea.ac.uk,t.hollowood@swansea.ac.uk,\\ s.p.kumar@swansea.ac.uk, andrea.legramandi@swansea.ac.uk,\\ n.talwar.2017429@swansea.ac.uk}
\abstract{
We consider the issue of information recovery for an object carrying energy and entropy into a black hole using the generalized entropy formalism, in the context of JT gravity where the backreaction problem can be solved exactly. We verify the main aspects of the Hayden-Preskill scenario but with some refinements. We show that the information is encoded in the Hawking radiation in a redundant way, as expected for a quantum error correcting code. We show how quantum extremal surfaces associated to information recovery have the form of a python's lunch and thereby show that the complexity of decoding is exponential in a combination of the entropy shift of the black hole and the entropy of the object. We also show that an infalling observer must have a smooth experience at the horizon and we calculate their endurance proper time inside the black hole before they are radiated out.
}

\begin{document}

\maketitle


\section{Introduction}

The fact that holography defines a quantum theory of gravity in spacetimes that are asymptotically AdS has led to spectacular progress in understanding quantum black holes \cite{Ryu:2006bv,Hubeny:2007xt,Faulkner:2013ana,Engelhardt:2014gca,Engelsoy:2016xyb,Penington:2019npb,Almheiri:2019psf,Jafferis:2015del}. In this new understanding, quantum mechanics is fundamental and gravity is an emergent phenomenon. It is now clear that black holes evaporate in a way that is perfectly consistent with quantum mechanics and, moreover, this can be seen even in the semi-classical limit \cite{Almheiri:2020cfm,Penington:2019kki,Almheiri:2019qdq,Goto:2020wnk,Marolf:2020rpm,Colin-Ellerin:2020mva}. The latter does not directly see the correlations within the Hawking radiation required to recover unitarity, but it has a subtle  way of performing the necessary book-keeping in the form of `entanglement islands' or `islands' for short \cite{Hollowood:2020cou,Hollowood:2020kvk,Hollowood:2021nlo,Hollowood:2021wkw,Hollowood:2021lsw,Bousso:2021sji,Wang:2021woy,Karananas:2020fwx,Basak:2020aaa,Matsuo:2020ypv,Hernandez:2020nem,Ling:2020laa,Chen:2020hmv,Chen:2020jvn,Chandrasekaran:2020qtn,Chen:2020uac,Hashimoto:2020cas,Gautason:2020tmk,Almheiri:2019hni,Kawabata:2021hac,Almheiri:2019yqk,Bhattacharya:2021jrn,Grimaldi:2022suv}. 

The semi-classical analysis can describe how information, in the form of a system carrying energy and entropy that falls into the black hole, can be recovered in the Hawking radiation. In other words, the information-theoretic analysis of Hayden and Preskill \cite{Hayden:2007cs} can be verified using only semi-classical methods, as in the pioneering work of Penington \cite{Penington:2019npb}. The purpose of the present work is to consider this in detail in the context of the JT gravity model \cite{Jackiw:1984je,Teitelboim:1983ux}, where the backreaction of the infalling system can be solved exactly, although one can expect a lot of the results to apply more universally.

Key to the recent progress in understanding quantum gravity at the semi-classical level is the generalized entropy, or Quantum Extremal Surface (QES), formalism. This gives a way to compute the entropy of the Hawking radiation $R$ emitted by the black hole up to some time as the solution of a variational problem over possible `islands' $I$, which is defined just in terms of quantities that can be calculated in the semi-classical regime
\EQ{
S(R)=\min_I S_I(R)\ ,\quad S_I(R)\equiv\underset{\partial I}\ext\Big\{\frac{\text{Area}(\partial I)}{4G_N}+S(\rho^\text{sc}_{R\cup I})\Big\}\ .
\label{hys}
} 
In the above, $I$ is a subregion of a Cauchy slice, containing $R$, that passes through the co-dimensional 2 QES $\partial I$ that are determined by the variational problem.\footnote{More precisely, the island is the domain of dependence of this subregion.} The second term is the entropy of the reduced state of the effective matter QFT on the subregion $R\cup I$ of the Cauchy slice, including the island $I$, defined on the fixed background metric of the black hole. We will refer to the different islands that can compete in the variational problem as `entropy saddles' since they are computed at the semi-classical level by saddles of the functional integral.

It is crucial to understand what the entropy $S(R)$ refers to. It is not necessarily equal to the entropy of the state of the radiation in the effective theory, the von Neumann entropy $S(\rho^\text{sc}_R)$. This is why Hawking's original conclusion for information loss is avoided. Rather, it is the entropy of the microscopic state of the radiation $\rho_R$ which is not the semi-classical state $\rho_R^\text{sc}$. These two states are associated to different Hilbert spaces, the fundamental one of quantum gravity (whatever that is), while the other is the space of excitations around a fixed background. It is remarkable that the QES formula allows us to calculate this microscopic entropy $S(R)\equiv S(\rho_R)$ even when we are ignorant of the details of the microscopic theory. 

So there are two levels of description at play, a microscopic one described by a quantum state $\rho$ and an effective description describing QFT over a fixed background geometry with a state $\rho^\text{sc}$.\footnote{The discussion could be couched in  more precise terms of operator algebras but it is simpler to talk about states even if it is only approximate.} The effective state is then embedded in the microscopic description via a linear map $V:\ \rho^\text{sc}\to\rho$, the `holographic map'. In the case where there is a holographic description, like for the black hole in JT gravity, the microscopic state is a state of the dual boundary non-gravitational theory. In such a scenario, there is clean split of the microscopic system into the boundary theory and the radiation bath, a half Minkowski space that is glued to the boundary of AdS$_2$ and is used to collect the Hawking radiation \cite{Almheiri:2019psf,Almheiri:2019yqk}, so the total microscopic Hilbert space factorizes ${\cal H}={\cal H}_B\otimes{\cal H}_R$. For a higher dimensional black hole in asymptotically flat space, ${\cal H}_R$ can be identified as describing the outgoing Hawking modes far from the black hole, effectively at $\mathscr I^+$\footnote{See, for example, \cite{Ghosh:2021axl}, for the application of the island formula in gravitating baths.}. 

It is important for our analysis that $R$ can also consist of disconnected subsets of the outgoing Hawking radiation. In this case, the island $I$ that dominates depends implicitly via the variational problem on the choice of $R$. This allows us to calculate the correlation between different subsets in the form of the mutual information $I(R_1,R_2)=S(R_1)+S(R_2)-S(R_1\cup R_2)$.

The interpretation of the QES formula \eqref{hys} is highly non-trivial. Firstly, the fact that the entropy of the microscopic state $\rho_R$ is not necessarily equal to the entropy of the effective state $\rho_R^\text{sc}$ is a consequence of the holographic map 
\EQ{
\rho=V\rho^\text{sc}V^\dagger\ .
}
In recent work \cite{AEHPV:22}, it has been argued that $V$ is a non-isometric map that acts trivially on the radiation. It is this fact that allows $\rho_R\neq\rho_R^\text{sc}$, as is implicit in the QES formula when the island is non-trivial. The QES formula implies that there is a `decoding map', an isometry that extracts the product state $\rho^{\text{sc}}_{R\cup I}\otimes\tilde \rho$ out of $\rho_R$, where the second factor accounts for the microscopic entropy that is responsible for the area term in \eqref{hys} \cite{Akers:2021fut}. 
This is clearer when the gravitational region also enjoys  a holographic description, since we can interpret the area term as the usual  Ryu-Takayanagi contribution to the entropy evaluated at the QES \cite{Faulkner:2013ana,Engelhardt:2014gca}.

The fact that the island, which is generally a behind-the-horizon region of the black hole, appears encoded in the microscopic state of the radiation, means physically that the quantum information inside the black hole has been evaporated out of the black hole. This process is non-local from the point-of-view of the effective theory as a result of the fact that $V$ is non-isometric \cite{AEHPV:22}.\footnote{The non-isometric map $V$ can be realized as an isometry along with post selection. The post-selection allows information to leak out of the black hole via quantum teleportation.}

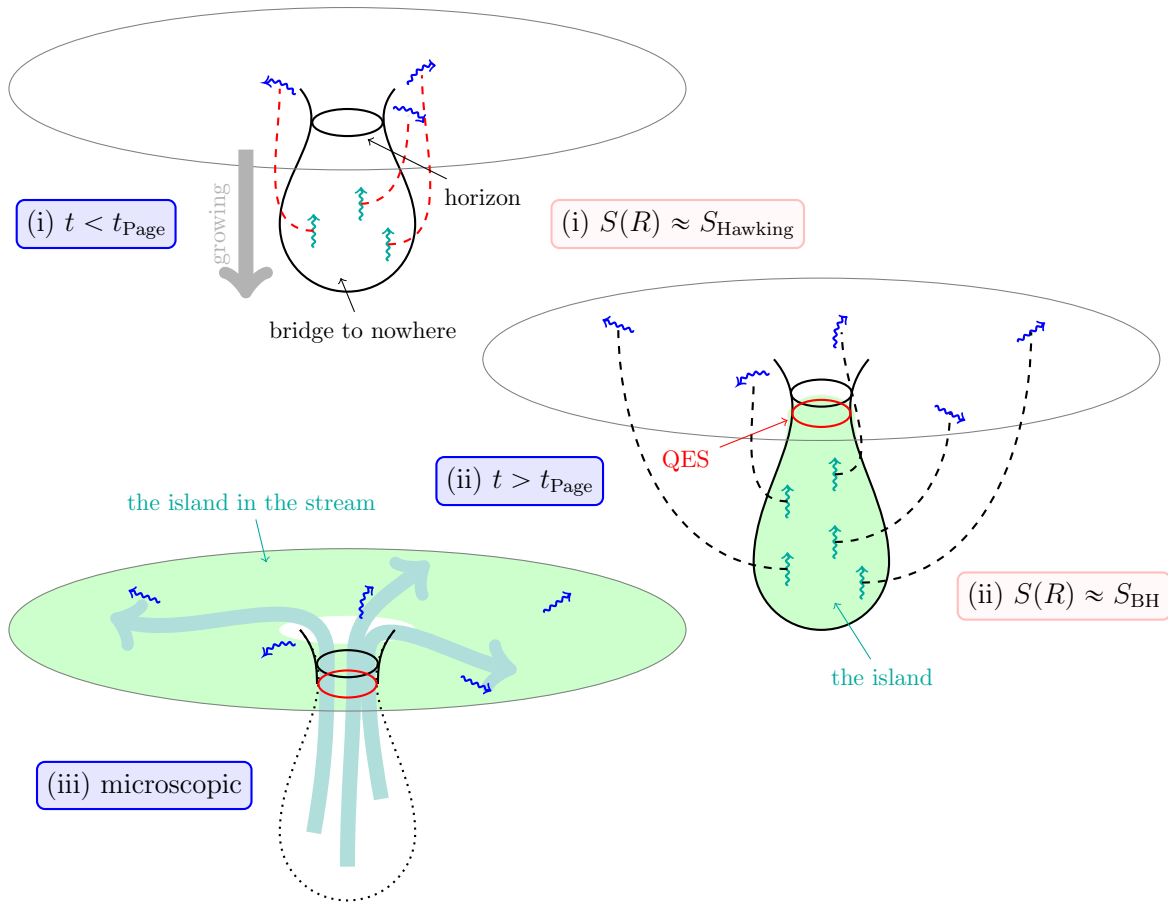
\begin{figure}[ht]
\begin{center}
\begin{tikzpicture}[scale=0.9, every node/.style={scale=0.9}]
\draw[thick] (0.3,0) to[out=-50,in=90] (0,-2) to[out=-90,in=180] (1,-3) to[out=0,in=-90] (2,-2) to[out=90,in=-130] (1.7,0);
\draw[black!50] (1,0) ellipse (5cm and 1.2cm);
\draw[thick] (1,-0.5) ellipse (0.52cm and 0.2cm);
\begin{scope}[xshift=2.1cm,yshift=0.2cm,rotate=35]
\draw[blue,decorate,thick,decoration={snake,segment length=1mm,amplitude=0.2mm},->] (-0.25,0) to[out=30,in=150] (0.25,0); 
\end{scope}
\begin{scope}[xshift=1.9cm,yshift=-0.4cm,rotate=-25]
\draw[blue,decorate,thick,decoration={snake,segment length=1mm,amplitude=0.2mm},->]  (-0.25,0) to[out=30,in=150] (0.25,0); 
\end{scope}
\begin{scope}[xshift=0cm,yshift=0cm,rotate=160]
\draw[blue,decorate,thick,decoration={snake,segment length=1mm,amplitude=0.2mm},->]  (-0.25,0) to[out=-30,in=-150] (0.25,0);  
\end{scope}
\begin{scope}[xshift=1.6cm,yshift=-2.3cm,rotate=90]
\draw[Emerald,decorate,thick,decoration={snake,segment length=1mm,amplitude=0.2mm},->]  (-0.25,0) -- (0.25,0);  
\end{scope}
\begin{scope}[xshift=1.2cm,yshift=-1.7cm,rotate=90]
\draw[Emerald,decorate,thick,decoration={snake,segment length=1mm,amplitude=0.2mm},->]  (-0.25,0) -- (0.25,0);  
\end{scope}
\begin{scope}[xshift=0.5cm,yshift=-2.1cm,rotate=90]
\draw[Emerald,decorate,thick,decoration={snake,segment length=1mm,amplitude=0.2mm},->]  (-0.25,0) -- (0.25,0); 
\end{scope}
\draw[thick,red,dashed] (0.5,-2.1) to[out=180,in=-90] (0,0);
\draw[thick,red,dashed] (1.2,-1.7) to[out=0,in=-90] (1.9,-0.4);
\draw[thick,red,dashed] (1.6,-2.3) to[out=0,in=-90] (2.1,0.2);
\node[right,fill=blue!10, draw=blue, thick, rounded corners=3pt] at (-3.9,-2) {(i) $t<t_\text{Page}$};
\node[right,fill=pink!10, draw=pink, thick, rounded corners=3pt] at (4,-2) {(i) $S(R)\approx S_\text{Hawking}$};
\draw[black!30,line width=2mm,<-] (-0.5,-3.1) -- (-0.5,-0.9) node[sloped,midway,above] {\footnotesize growing};
\node[right] (a2) at (-0.3,-3.6) {\footnotesize bridge to nowhere};
\draw[->] (a2) -- (0.9,-2.8);
\node (a1) at (3,-1.6) {\footnotesize horizon};
\draw[->] (a1) -- (1.3,-0.75);
\begin{scope}[xshift=7cm,yshift=-4cm]
\draw[pink!20,fill=green!20] (0.5,-0.8) to[out=-90,in=90] (0,-3) to[out=-90,in=180] (1,-4) to[out=0,in=-90] (2,-3) to[out=90,in=-90] (1.5,-0.8) to[out=110,in=70] (0.5,-0.8);
\draw[thick] (0.3,0) to[out=-50,in=90] (0,-3) to[out=-90,in=180] (1,-4) to[out=0,in=-90] (2,-3) to[out=90,in=-130] (1.7,0);
\draw[black!50] (1,0) ellipse (5cm and 1.2cm);
\draw[thick] (1,-0.5) ellipse (0.45cm and 0.2cm);
\draw[thick,red] (1,-0.8) ellipse (0.43cm and 0.2cm);
\begin{scope}[xshift=4.1cm,yshift=0.4cm,rotate=35]
\draw[blue,decorate,thick,decoration={snake,segment length=1mm,amplitude=0.2mm},->] (-0.25,0) -- (0.25,0); 
\end{scope}
\begin{scope}[xshift=2.9cm,yshift=-0.8cm,rotate=-25]
\draw[blue,decorate,thick,decoration={snake,segment length=1mm,amplitude=0.2mm},->]  (-0.25,0) -- (0.25,0); 
\end{scope}
\begin{scope}[xshift=-2cm,yshift=0.5cm,rotate=160]
\draw[blue,decorate,thick,decoration={snake,segment length=1mm,amplitude=0.2mm},->]  (-0.25,0) -- (0.25,0);  
\end{scope}
\begin{scope}[xshift=1.3cm,yshift=0.4cm,rotate=70]
\draw[blue,decorate,thick,decoration={snake,segment length=1mm,amplitude=0.2mm},->]  (-0.25,0) to[out=30,in=150] (0.25,0); 
\end{scope}
\begin{scope}[xshift=0cm,yshift=-0.3cm,rotate=-160]
\draw[blue,decorate,thick,decoration={snake,segment length=1mm,amplitude=0.2mm},->]  (-0.25,0) to[out=-30,in=-150] (0.25,0);  
\end{scope}
\begin{scope}[xshift=1.2cm,yshift=-1.7cm,rotate=90]
\draw[Emerald,decorate,thick,decoration={snake,segment length=1mm,amplitude=0.2mm},->]  (-0.25,0) -- (0.25,0);  
\end{scope}
\begin{scope}[xshift=0.5cm,yshift=-2.1cm,rotate=90]
\draw[Emerald,decorate,thick,decoration={snake,segment length=1mm,amplitude=0.2mm},->]  (-0.25,0) -- (0.25,0); 
\end{scope}
\begin{scope}[yshift=-1cm]
\begin{scope}[xshift=1.6cm,yshift=-2.3cm,rotate=90]
\draw[Emerald,decorate,thick,decoration={snake,segment length=1mm,amplitude=0.2mm},->]  (-0.25,0) -- (0.25,0);  
\end{scope}
\begin{scope}[xshift=1.2cm,yshift=-1.7cm,rotate=90]
\draw[Emerald,decorate,thick,decoration={snake,segment length=1mm,amplitude=0.2mm},->]  (-0.25,0) -- (0.25,0);  
\end{scope}
\begin{scope}[xshift=0.5cm,yshift=-2.1cm,rotate=90]
\draw[Emerald,decorate,thick,decoration={snake,segment length=1mm,amplitude=0.2mm},->]  (-0.25,0) -- (0.25,0); 
\end{scope}
\end{scope}
\draw[thick,dashed] (0.5,-3.1) to[out=180,in=-90] (-2,0.5);
\draw[thick,dashed] (1.2,-2.7) to[out=0,in=-90] (2.9,-0.8);
\draw[thick,dashed] (1.6,-3.3) to[out=0,in=-90] (4.1,0.4);
\draw[thick,dashed] (1.2,-1.7) to[out=0,in=-90] (1.3,0.4);
\draw[thick,dashed] (0.5,-2.1) to[out=180,in=-90] (0,-0.3);
\node[right,fill=blue!10, draw=blue, thick, rounded corners=3pt] at (-4.7,-1.8) {(ii) $t>t_\text{Page}$};
\node[right,fill=pink!10, draw=pink, thick, rounded corners=3pt] at (3,-3.5) {(ii) $S(R)\approx S_\text{BH}$};
\node[red] (b1) at (-1,-1.5) {\footnotesize QES};
\draw[red,->] (b1) -- (0.5,-0.9);
\node[Emerald,right] (a2) at (1,-4.7) {\footnotesize the island};
\draw[Emerald,->] (a2) -- (1.2,-3.6);
\end{scope}
\begin{scope}[xshift=0cm,yshift=-8cm]
\draw[green!20,fill=green!20] (1,0) ellipse (5cm and 1.2cm);
\draw[white,fill=white] (1,0) ellipse (1cm and 0.2cm);
\draw[Emerald!30,line width=2mm,->] (0.5,-3) to[out=80,in=-90] (0.7,-0.4) to[out=90,in=-20] (-2.5,0.2);
\draw[Emerald!30,line width=2mm,->] (1,-3.5) to[out=90,in=-90] (1.1,-0.4) to[out=90,in=-140] (2.2,1);
\draw[Emerald!30,line width=2mm,->] (1.5,-2.5) to[out=100,in=-90] (1.3,-0.4) to[out=90,in=170] (3.5,-0.6);
\draw[thick] (0.3,0) to[out=-50,in=90] (0.55,-0.8);
\draw[thick] (1.7,0) to[out=-130,in=90] (1.45,-0.8);
\draw[thick,dotted] (0.3,0) to[out=-50,in=90] (0,-3) to[out=-90,in=180] (1,-4) to[out=0,in=-90] (2,-3) to[out=90,in=-130] (1.7,0);
%
\draw[black!50] (1,0) ellipse (5cm and 1.2cm);
\draw[thick] (1,-0.5) ellipse (0.45cm and 0.2cm);
\draw[thick,red] (1,-0.8) ellipse (0.43cm and 0.2cm);
\begin{scope}[xshift=4.1cm,yshift=0.4cm,rotate=35]
\draw[blue,decorate,thick,decoration={snake,segment length=1mm,amplitude=0.2mm},->] (-0.25,0) -- (0.25,0); 
\end{scope}
\begin{scope}[xshift=2.9cm,yshift=-0.8cm,rotate=-25]
\draw[blue,decorate,thick,decoration={snake,segment length=1mm,amplitude=0.2mm},->]  (-0.25,0) -- (0.25,0); 
\end{scope}
\begin{scope}[xshift=-2cm,yshift=0.5cm,rotate=160]
\draw[blue,decorate,thick,decoration={snake,segment length=1mm,amplitude=0.2mm},->]  (-0.25,0) -- (0.25,0);  
\end{scope}
\begin{scope}[xshift=1.3cm,yshift=0.4cm,rotate=70]
\draw[blue,decorate,thick,decoration={snake,segment length=1mm,amplitude=0.2mm},->]  (-0.25,0) to[out=30,in=150] (0.25,0); 
\end{scope}
\begin{scope}[xshift=0cm,yshift=-0.3cm,rotate=-160]
\draw[blue,decorate,thick,decoration={snake,segment length=1mm,amplitude=0.2mm},->]  (-0.25,0) to[out=-30,in=-150] (0.25,0);  
\end{scope}
\node[Emerald,right] (a2) at (-2.4,1.9) {\footnotesize the island in the stream};
\draw[Emerald,->] (a2) -- (-0.2,1);
\node[right,fill=blue!10, draw=blue, thick, rounded corners=3pt] at (-3.6,-2.3) {(iii) microscopic};
\end{scope}
\end{tikzpicture}
\caption{\footnotesize The evolution of entanglement as the black hole evaporates. At early time (i) Hawking modes and their entangled partners behind the horizon are pair produced and the ER `bridge to nowhere' grows from the horizon. The entropy of the radiation is the thermal entropy, as calculated by Hawking. After the Page time (ii), a QES develops just behind the horizon and inside this is the island. This is the semi-classical way of accounting for the fact that the island has been scrambled and evaporated out in the radiation---the `island in the stream'---at the microscopic level as suggested in (iii). The entanglement between the Hawking modes and their partners is now cancelled and the entropy of the radiation is proportional to the area of the QES, approximately $S_\text{BH}$.}
\label{fig1} 
\end{center}
\end{figure}

The island formalism described above gives the following description of an evaporating black hole. The map $V$ undergoes a qualitative change at the Page time \cite{Page:1993wv,Page:1993df} when the entropy becomes dominated by an island saddle. In figure \ref{fig1} we have a schematic picture of the entanglement structure of an evaporating black hole. Before the Page time, Hawking modes are produced as entangled partners either side of the horizon \cite{Hawking:1974sw} and the inside of the black hole grows. This is the `bridge to nowhere' of \cite{Brown:2019rox} shown in (i). In this regime there is no island and $\rho_R\approx \rho^\text{sc}_R$, so the entropy of the radiation is just the thermal entropy of the radiation, up to a UV divergence, as calculated by Hawking \cite{Hawking:1976ra}. At the Page time, in (ii), there is an almost instantaneous change in the map $V$ which affects the relation between semiclassical and fundamental radiation state, $\rho_R\not\approx\rho^\text{sc}_R$; this is signaled by the generalized entropy being dominated by an island saddle. The bridge to nowhere up to the QES, which lies just inside the horizon, becomes the island. This is the way that the effective theory accounts for the fact that the quantum state of the Hawking partners in the island have been scrambled up and evaporated out of the black hole as represented in (iii). The radiation now includes both the Hawking modes and their partners and so their entanglement no longer contributes to the entropy of the radiation. The latter now comes from the area of the QES that divides the radiation and the black hole, approximately the Bekenstein-Hawking entropy of the black hole $S_\text{BH}$.

\subsection{The Hayden-Preskill scenario}

The purpose of this work is to consider what happens when an auxiliary system $D$ carrying energy and entropy, a diary, to use the terminology of Hayden and Preskill \cite{Hayden:2007cs}, falls into a black hole. 
Information recovery is a key part of the analysis, so it might be useful to quickly review this. Suppose we have some quantum system $D$ which is added to some other system to make a total system $B_0$ which then undergoes some time evolution that mixes up the systems. We want to know what it means to recover the information of $D$ in some subsystem $R\subset B_0$. For us $R$ will be the radiation emitted by the black hole up to a certain time and the complement $B$ is the remaining black hole, so $B_0=B\cup R$. In general, the initial state of $D$ can be mixed and it is a standard trick to introduce an auxiliary system $\overline D$ that purifies it as a book-keeping device so $S(\overline D)=S(D)$.

Since the overall quantum state on $R\cup B\cup\overline D$ is pure, the entropies of the 3 subsystems must form the sides of a triangle. This follows from subadditivity and the Araki-Lieb inequality. The condition that $D$ can be recovered in $R$ is that the triangle degenerates in the form:
\begin{center}
\begin{tikzpicture} [scale=0.6]
\draw[very thick] (0,0) -- (4,4) node[sloped,midway,above] {$S(B)$} -- (6,0) node[sloped,midway,above] {$S(D)$} -- (0,0) node[sloped,midway,below] {$S(R)$};
\draw[line width=1mm,->] (8,2) -- (12,2) node[sloped,midway,above] {\footnotesize recovery};
\begin{scope}[xshift=14cm,yshift=1.5cm]
\draw[very thick] (0,0.2) -- (4.5,0.2) node[sloped,midway,above] {$S(B)$};
\draw[very thick] (4.6,0.2)  -- (6,0.2) node[sloped,midway,above] {$S(D)$};
\draw[very thick] (0,0) -- (6,0) node[sloped,midway,below] {$S(R)$};
\end{scope}
\end{tikzpicture}
\end{center}
so that we get maximal mutual information, i.e.~maximal entanglement, between $\overline{D}$ and $R$:
\EQ{
S(R)=S(D)+S(B)\quad\implies\quad I(\overline D,R)=2S(D)\quad \text{and}\quad I(\overline D,B)=0\ .
\label{pcc}
}
Note that this implies that the reduced state $\rho_{\overline DB}=\rho_{\overline D}\otimes\rho_B$ completely factorizes. 

Quantum information is more subtle than its classical counterpart in the sense that the information in $D$ can also be recovered in other tensor factors of the Hilbert space, a feature that will be important in our analysis. This kind of redundancy is exploited by Quantum Error Correcting Codes (QECC) and lies behind the robustness of these codes.

We will present three levels of analysis of information recovery, using the QES formula \eqref{hys} to compute the relevant entropies. The variational problem becomes tractable in the case when the black hole is evaporating slowly, the quasi-adiabatic regime, which is true for most of the black hole's life. This approximation is used in Hawking's analysis, where it is meaningful to associate a slowly varying temperature $T$ to the black hole, and only breaks down near the end of the evaporation. 
In the first level, we assume that $D$ is essentially a shockwave, i.e.~a very narrow pulse of energy and entropy, and the time scales involved in recovering the information are large compared with the thermal time scale $ T^{-1}$ and also the longer scrambling time of the black hole $T^{-1}\log S_\text{BH}/\cc$, where $\cc$ is the number of free massless fields propagating in the effective QFT in the black hole background. In the second level of analysis, we consider effects that are of the order of the scrambling time,  leading to refinements that are logarithmic in the entropies. In the final level of analysis, we model $D$ as a finite width pulse. This allows us to investigate numerically the so-called python's lunch associated to $D$ which is conjectured to give a measure of the difficulty of decoding $D$ once sufficient information has been evaporated out of the black hole into the radiation \cite{Brown:2019rox,Engelhardt:2021qjs}.

The questions to be addressed are:
\begin{enumerate}[label=\protect\circled{\arabic*}]
\item How does $D$ affect the transition at the Page time?
\item When is the information contained in $D$ returned to the outside in the radiation (this is the Hayden-Preskill scenario \cite{Hayden:2007cs}).
\item Can an arbitrarily complicated process on the radiation affect $D$ inside the black hole?
\item What is the `experience' of $D$ given that their information is to be scrambled and evaporated out of the black hole eventually? How much of the interior can $D$ explore?
\item How does the consideration of the backreaction of $D$ affect the $A=R_B$ scenario which involves an observer, i.e.~$D$, falling into the black hole to verify that there is a smooth horizon and no firewall.
\item Can the information in $D$ be reconstructed in other subsets of the Hawking radiation and, in particular, in the late radiation emitted after $D$ falls in?
\end{enumerate}

The paper is organized as follows. In section \ref{s2} we describe the generalized entropy formalism when the black hole is evaporating slowly, the adiabatic regime, and when there are also infalling objects, essentially shockwaves, carrying energy and entropy. This allows us to describe the process of information recovery in terms of an exchange of saddles of the generalized entropy. The main elements of the Hayden-Preskill analysis \cite{Hayden:2007cs} follow in a simple way. Information is only recovered after the Page time and after a delay that Hayden and Preskill identified as a scrambling time for the black hole. Here, we will find that the time delay is modified by the backreaction of the object. The details of the computations on which this section is based, and in particular the extremization of the generalized entropy in the shockwave geometry for an arbitrary number of radiation intervals and islands, are presented in appendix \ref{app:A}.
In section \ref{s4} we consider the information recovery more precisely and show that the result is subject to certain corrections logarithmic in the entropy of the black hole and $D$. In the analysis we model the object as infinitely narrow in time but relax this in section \ref{s5} in order to see how the generalized entropy behaves for a smooth object. This shows that there exists a new extremum of the generalized entropy which has a higher entropy than the two extrema that can dominate the entropy. This is precisely an example of a python's lunch which is known to be related to the difficulty of decoding, in this case, the information of $D$ in the radiation \cite{Brown:2019rox,Engelhardt:2021qjs}. In section \ref{sec:6} we discuss how
the backreaction of the infalling observer destroys the entanglement between a newly emitted Hawking quantum and the old radiation, allowing for a smooth experience across the horizon.
However, the infalling observer must be, at a certain point, scrambled and evaporated out. We then compute the endurance proper time inside the black hole for an ultra-relativistic diary. We conclude in section \ref{s7} with a summary of the main results.

\section{Islands-in-the-stream with infalling objects}\label{s2}

The calculations in this section are done in the JT gravity model, however, we write the results in a way which does not depend on the details of the model and, by analogy, can be applied to higher-dimensional black holes, including Schwarzschild, in the $s$-wave approximation that dominates Hawking emission.

\subsection{Review of islands-in-the-stream without infalling objects}

For most of its life a black hole evaporates slowly and the adiabatic, or quasi-static, approximation applies. Specifically this is in the limit $S_\text{BH}\gg c$ and it is meaningful to associate a slowly varying temperature $T(u)$ to the black hole as a function of the outgoing null Eddington-Finkelstein coordinate $u$ on $\mathscr I^+$.\footnote{For a higher dimensional black hole the geometry outside the zone is described by an outgoing Vaidya metric. For a black hole in JT gravity the situation is simpler since the geometry is fixed to be AdS$_2$.}
If we collect the radiation in a set of large intervals $R = \bigcup_j [u_{2j-1},u_{2j}]\subset\mathscr I^+$, then the entropy is just the thermodynamic entropy of a relativistic gas of $\cc$ species of particles contained in $R$:
\EQ{
S_\text{rad}(R)\approx\frac{\pi \cc}{6}\int_R T(u)\,du\ .
\label{tat}
}
`Large' in this context requires that the entropy of each sub-interval in $R$ and its complement is much larger than $\cc$, specifically we require $|u_j-u_{j+1}| \gg 1/T(u_j)$. In \eqref{tat} $\approx$ means equality within the scope of the adiabatic approximation. The above expression assumes that we have regularized the UV divergences that appear in the entropy of a subregion in a QFT. Ultimately we are interested in entropy differences which are in any case UV safe.\footnote{In the following, we often consider adjoining subsets and assume $S_\text{rad}(A\cup B)=S_\text{rad}(A)+S_\text{rad}(B)$. Implicitly, we assume a small gap between such sets that is bigger than the UV cut off.}

Equation \eqref{tat} gives the entropy of the no-island $I=\varnothing$ solution of the variational problem \eqref{hys}. If we want to consider island saddles, we have to extremize the generalized entropy \eqref{hys}. The extremization for a generic number of intervals was performed in \cite{Hollowood:2021nlo}. In order to state the result, it is useful to introduce Kruskal-Szekeres (KS) null coordinates $(U,V)$ which are related to the Eddington-Finkelstein outgoing/ingoing coordinates $(u=t-r_*,v=t+r_*)$ via exponential maps
\EQ{
U=-e^{-\sigma(u)}\ ,\qquad V=e^{\sigma(v)}\  ,
\label{ruz}
}
where the function $\sigma(t)$ in the adiabatic limit is related to the slowly varying temperature
\EQ{
\frac{d\sigma}{dt}=2\pi T\ .
}
The extremization gives the following results:
\begin{enumerate}[label=\protect\circled{\arabic*}]
  \item In the adiabatic limit, all the QES are very close to and inside the horizon, with KS coordinates satisfying\footnote{Here, we are assuming that the extremal entropy $S_*$ is negligible, otherwise one replaces $S_\text{BH}(v)$ by $S_\text{BH}(v)-S_*$.}
\EQ{
UV\approx\frac{\cc}{48S_\text{BH}(v)}\ll1\ .
\label{yet}
}
In the above, $S_\text{BH}(v)$ is the instantaneous Bekenstein-Hawking entropy of the black hole as a function of the infalling coordinate $v$ of the QES, with the area interpreted as the value of the dilaton in JT gravity.

\item The possible islands $I$ are the domain of dependence of a set of intervals behind the horizon on a Cauchy slice that includes $R$ and the QES, the boundary $\partial I$, whose reflection in the horizon $U\to-U$ denoted $I\to\tilde I$, are such that $u_{\partial\tilde I}\subset\partial R$.
\end{enumerate}
We call the image $\tilde I$ the `island in the stream' because it shows exactly where in the Hawking radiation $R$ the quantum information of modes in the island have been scrambled and evaporated out of the black hole and are available for decoding on $\mathscr I^+$. The infalling $v$ coordinate of a QES is then related to the associated endpoint $u_j \in \partial R$ on $\mathscr I^+$:
\EQ{
v_{\partial I}=u_j-\frac1{2\pi T(u_j)}\log\frac{S_\text{BH}(u_j)}\cc+\cdots\ ,
\label{sum}
}
For now, we will assume that the log term is subleading, although later and specifically in section \ref{s4}  we will take account of these corrections carefully because they are interpreted as the scrambling time of the black hole.

It is now simple to calculate the various term in the generalized entropy. Since $I$ contains the purifiers of the Hawking modes, the modes in $I$ add to the matter field entropy $S(\rho^\text{sc}_{R\cup I})$ unless they are in the intersection $R\cap\tilde I$ on $\mathscr I^+$ in which case the entropy cancels between modes and purifiers. Hence, we can write  
\EQ{
S(\rho^\text{sc}_{R\cup I})\approx S_\text{rad}(R\ominus\tilde I)\ ,
}
in terms of the symmetric difference $R\ominus\tilde I=R\cup\tilde I-R\cap\tilde I$. 
On the other hand, the area terms contribute as the instantaneous entropy of the black hole since the QES is very close to the horizon, yielding
\EQ{
S_I(R)\approx\sum_{\partial\tilde I}S_\text{BH}(u_{\partial\tilde I})+S_\text{rad}(R\ominus\tilde I)\ .
\label{twm}
}
The variational problem then determines that the endpoints of the island-in-the-stream $\partial\tilde I\subset\partial R$ to leading order.

A more accurate treatment of the problem would require introducing the greybody factor to take into account that the Hawking modes must tunnel through a potential barrier in order to reach $\mathscr I^+$. Following \cite{Hollowood:2021lsw}, we are restricting to the case where the greybody factor just depends on the frequency and the black-hole temperature, which is only true for the Schwarzschild case, where there is a single scale in the problem (the Schwarzschild radius), or JT gravity, where the greybody factor can be introduced by hand as a semi-reflective barrier between the gravity region and the bath. Given this assumption, the greybody factor is simply described by an overall coefficient $\xi$ which controls the reversibility of the problem; in particular we have the following relation between the entropy flux of the black hole and the radiation:
\begin{equation}
    \frac{dS_\text{rad}}{dt}=-\xi \frac{dS_\text{BH}}{dt}\ ,
    \label{eq:irr}
\end{equation}
where $\xi\to1$ is the opaque limit which is the reversible case where evaporation is infinitely slow, while $\xi=2$ is the completely transparent case with no back-scattering of Hawking modes. In this section we leave $1<\xi\leq2$ arbitrary. Note that integrating \eqref{eq:irr} on a interval we find
\EQ{
S_\text{rad}(u_1,u_2)=\xi\big(S_\text{BH}(u_1)-S_\text{BH}(u_2)\big)\ ,
}
so that we can write \eqref{twm} entirely in terms of the Bekenstein-Hawking entropy.

\subsection{Adding infalling objects}
\label{sec:IITS_with_D}

It is possible to generalize these rules to include quantum objects $D$ that fall into the hole. We model $D$ as a narrow pulse of energy and entropy with a constant energy density in the time interval $t\in[t_0,t_0+\epsilon]$ on the AdS/Minkowski interface. Later in section \ref{s5} we will resolve what happens for finite $\epsilon$ interval but for now we work in the limit $\epsilon\to0$.

In the following, it will be important to recognise that, as well as carrying entropy into the black hole, $D$ also carries energy which results in a non-trivial backreaction on the geometry which will be crucial to include. The fact that $D$ must also carry energy can be seen as a consequence of the Generalized Second Law (GSL) of black hole thermodynamics \cite{Bekenstein:1972tm} which says that when a system, here $D$, carrying entropy $S(D)$, falls into a black hole then the backreaction is such that the change in the black hole's entropy satisfies
\EQ{
\Delta S_\text{BH}>S(D)\ .
\label{xxo}
}
Here, $\Delta S_\text{BH}=S_\text{BH}(t_0+\epsilon)-S_\text{BH}(t_0)$ is the jump in the black hole entropy caused by $D$. This changes the rate of entropy and energy flux in the Hawking radiation. We will assume that the jump in the entropy of the hole is small compared with its entropy
\EQ{
\Delta S_\text{BH}\ll S_\text{BH}(t_0)\ ,
\label{hsu}
}
but large in the sense that it must be bigger than the radiation entropy emitted in a thermal time
\EQ{
\Delta S_\text{BH}\gtrsim \cc\gg1\ .
\label{rvv}
}
This ensures that within the semi-classical approximation we are working in, $D$ is big enough to have interesting effects. 

One concrete way to model $D$ is as a smooth but narrow pulse created by a scalar primary operator quench in the matter theory with an operator of conformal dimension $h$. These pulses carry an energy that scales like
\EQ{
{\cal E}_D=\frac h\epsilon\ .
}
Hence, the backreaction produces 
\EQ{
\Delta S_\text{BH}\thicksim \frac{{\cal E}_D}{ T(t_0)}
\ .
\label{eq:1st_law}
}
Casini's version \cite{Casini:2008cr} of the Bekenstein bound \cite{Bekenstein:1980jp} implies an inequality between $\cal{E}_D$ and $S_D$ which can be used to show that the GSL is comfortably satisfied for this example.

In addition to the backreaction caused by the energy of $D$ there is also an entropy backreaction. This is simple to account for in the generalized entropy formalism. If $D$ lies in the island according to its infalling coordinate, i.e.~$v=t_0$, then its entropy contributes positively to $S_I(R)$ but negatively to $S_I(R\cup\overline D)$:
\EQ{
S_I(R)=\cdots+\delta_{I,D}S(D)\ ,\qquad S_I(R\cup \overline D)=\cdots+S(D)-\delta_{I,D}S(D)\ ,
\label{trr}
}
where $\delta_{I,D}=0$ or 1 when $D$ lies outside or inside the island $I$ according to its infalling coordinate.

We now describe how the islands-in-the-stream formalism is generalized to account for $D$. The detailed proofs appear in appendix \ref{app:A}. In the formalism, the QES $\partial I$ have an outgoing coordinate $u$ that equals one of the end-points of the radiation region $R$ at $\mathscr I^+$. When the infalling object $D$ is included, the entropy $S_\text{BH}(t)$ must now include the backreaction of $D$, i.e.~jumps discontinuously by $\Delta S_\text{BH}$ at $t=t_0$. In addition, there is a possible new QES just before $D$ falls in, i.e.~with infalling coordinate $v=t_0$ and mirror coordinate $u=t_0$ on $\mathscr I^+$. The more refined analysis in section \ref{s4} will give a more precise understanding of where this new QES lies. Hence, for the outgoing coordinates of the QES we have an enlarged set of possibilities $u_{\partial\tilde I}\subset \partial R\cup\{t_0\}$. 

We can use the diagrammatic representation of the possible islands introduced in \cite{Hollowood:2021nlo} in order to compute  when the information in $D$ is scrambled and evaporated out of the black hole and available for decoding in the Hawking radiation. The diagrams show the regions of $R$ at $\mathscr I^+$ along with the island in the stream $\tilde I$ and the symmetric difference $R\ominus\tilde I$.

\subsection{Information recovery in the early radiation}
\label{sec:HP_IITS}

The first scenario we consider is the one described by Hayden and Preskill \cite{Hayden:2007cs} where the radiation is collected from the beginning of the black hole at $t=0$, i.e.~$R=[0,u]$. The question is what is the minimum time needed to recover the information of $D$ in $R$? In other words, what is the minimal value for $u$ in order that $I(R,\overline D)=2S(D)$? 

Note that we must have at least $u
>t_0$ and consequently we split the radiation into two sets, $R=R_1\cup R_2$ with  $R_1=[0,t_0]$ and $R_2=[t_0,u]$. Both $S(R)$ and $S(R\cup\overline D)$ that are needed to compute $I(R,\overline D)$ each of which involve a competition of 3 saddles:
\EQ{
S(R)&=\min\big(S_\text{rad}(R),S_\text{BH}(t_0)+S_\text{rad}(R_2),S_\text{BH}(u)+S(D)\big)\ ,\\[5pt]
S(R\cup\overline D)&=\min\big(S_\text{rad}(R)+S(D),S_\text{BH}(t_0)+S_\text{rad}(R_2)+S(D),S_\text{BH}(u)\big)\ ,
\label{pop}
}
corresponding to the islands-in-the-stream $\tilde I=\varnothing$, $R_1$ and $R$, respectively. It is clear that information recovery $I(R,\overline D)=2S(D)$ requires that the island saddle $\tilde I_2=[0,u]$ dominates in both cases:
\begin{center}
\begin{tikzpicture} [scale=0.8]
\filldraw[fill = Plum!10!white, draw = Plum!10!white, rounded corners = 0.2cm] (-2.5,1.1) rectangle (6.6,-3.3);
%
\draw[decorate,very thick,black!40,decoration={snake,amplitude=0.03cm}] (0,-2.5) -- (0,0.5);
\draw[dotted,thick] (0,0) -- (6,0);
\draw[dotted,thick,red] (0,-1) -- (6,-1);
\draw[dotted,thick,blue] (0,-2) -- (6,-2);
\draw[dotted,thick,OliveGreen] (3,-2.5) -- (3,0.5);
\draw[dotted,thick] (4,-2.5) -- (4,0.5);
\draw[very thick] (4,0) -- (0,0);
\filldraw[black] (0,0) circle (2pt);
\filldraw[black] (4,0) circle (2pt);
	\draw[very thick,red] (2.7,-1) -- (-0.2,-1);
	\filldraw[red] (2.7,-1) circle (4pt);
	\filldraw[red] (-0.2,-1) circle (4pt);
	\filldraw[OliveGreen] (3,-1) circle (2.5pt);
	\draw[OliveGreen] (3,-1) circle (4pt);
	\draw[very thick,blue] (3,-2) -- (4,-2);
	\filldraw[blue] (3,-2) circle (2pt);
	\filldraw[blue] (4,-2) circle (2pt);
	\node at (4,-2.8) {\footnotesize$u$};
	\node at (0,-2.8) {\footnotesize$0$};
	\node[right] at (-2.5,0) {$R$};
	\node[right] at (-2.5,-1) {$\tilde I_1$};
	\node[right] at (-2.5,-2) {$R \ominus\tilde I_1$};	
	\node at (1.5,0.4) {\footnotesize$R_1$};
	\node at (3.5,0.4) {\footnotesize$R_2$};
	\node at (3,-2.8) {\footnotesize $t_0$};
	\begin{scope}[xshift=10cm]
\filldraw[fill = Plum!10!white, draw = Plum!10!white, rounded corners = 0.2cm] (-2.5,1.1) rectangle (6.6,-3.3);
%
	\draw[decorate,very thick,black!40,decoration={snake,amplitude=0.03cm}] (0,-2.5) -- (0,0.5);
	\draw[dotted,thick] (0,0) -- (6,0);
	\draw[dotted,thick,red] (0,-1) -- (6,-1);
	\draw[dotted,thick,blue] (0,-2) -- (6,-2);
	\draw[dotted,thick,OliveGreen] (3,-2.5) -- (3,0.5);	
	\draw[very thick] (4,0) -- (0,0);
	\filldraw[black] (0,0) circle (2pt);
	\filldraw[black] (4,0) circle (2pt);
	\draw[very thick,red] (4,-1) -- (-0.2,-1);
	\filldraw[red] (4,-1) circle (4pt);
	\filldraw[red] (-0.2,-1) circle (4pt);
	\filldraw[OliveGreen] (3,-1) circle (2.5pt);
	\draw[OliveGreen] (3,-1) circle (4pt);
		%
	%
	\node[right] at (-2.5,0) {$R$};
	\node[right] at (-2.5,-1) {$\tilde I_2$};
	\node[right] at (-2.5,-2) {$R \ominus\tilde I_2$};	
		\end{scope}
	\end{tikzpicture}
\end{center}
\noindent In these diagrams the red blob on the left is the QES of the extremal black hole, which contributes the extremal entropy $S_*$, which existed before the black hole was excited by some infalling matter at $t=0$. We will take $S_*$ to be negligible.\footnote{Alternatively, in a higher dimensional black holes, the left blob might not be a QES but it can indicate the origin of the polar coordinates.} The green circle is the diary. So recovery requires
\EQ{
S_\text{BH}(u)+S(D)<\min\big(S_\text{rad}(R),S_\text{BH}(t_0)+S_\text{rad}(R_2)\big)\ .
}

Hence, if $D$ falls in before the Page time, i.e.~$S_\text{rad}(R_1)<S_\text{BH}(t_0)$, then recovery involves a direct transition between saddles with islands $\varnothing\to I_2$. This happens first for $S(R\cup\overline D)$ when
\EQ{
S_\text{rad}(R)=S_\text{BH}(u)-S(D)\ ,
}
and signals the fact that $I(R,\overline D)$ begins to rise. Note that this happens at a time which is just before the Page time of the black hole $S_\text{BH}(R)=S_\text{BH}(u)$. Recovery is then completed when $S(R)$ makes the transition which is just after the Page time. So the information comes out continuously as $R$ increases across the interval
\EQ{
\frac\xi{\xi+1}\big(S_\text{BH}(t_0)-S_\text{rad}(R_1)+&\Delta S_\text{BH}-S(D)\big)\leq S_\text{rad}(R_2)\\[5pt] &\leq 
\frac\xi{\xi+1}\big(S_\text{BH}(t_0)-S_\text{rad}(R_1)+\Delta S_\text{BH}+S(D)\big)\ .
}
So, if $D$ falls in early, one will have to wait till after the Page time in order to recover the information. Moreover, the information comes out continuously between the saddle transition for $R\cup\overline D$ and $R$. This will emerge as a universal feature for information recovery.

On the other hand, if $D$ falls in after the Page time, there are transitions $\varnothing\to I_1\to I_2$, the former being the transition at the Page time familiar in case with no $D$. Before the second transition $I_1\to I_2$, there is no information of $D$ in the radiation $I(R,\overline D)=0$. The transition happens first for $S(R\cup\overline D)$ when $I(R,\overline D)$ starts to increase proportional to $S_\text{rad}(R_2)$ until $S(R)$ also makes the transition $I_1\to I_2$ and all the information is recovered. Hence, the mutual information is always continuous:
\EQ{
I(R,\overline D)=\max\Big(0\,,\,\frac{\xi+1}\xi S_\text{rad}(R_2)+S(D)-\Delta S_\text{BH}\, ,\,2S(D)\Big)\ ,
\label{fuz}
}
meaning that after a hiatus the information comes out continuously until it is all out in the interval where $R_2$ increases subject to
\EQ{
\frac\xi{\xi+1}\big(\Delta S_\text{BH}-S(D)\big)\leq S_\text{rad}(R_2)\leq \frac\xi{\xi+1}\big(\Delta S_\text{BH}+S(D)\big)\ .
}
These expressions show clearly the important r\^ole that the GSL plays in the analysis. 

Let us call the radiation after $D$ falls in, when the mutual information vanishes, $R_2=R^\natural$, i.e.~$I(R_1\cup R^\natural,\overline D)=0$. We have
\EQ{
S_\text{rad}(R^\natural)=\frac\xi{\xi+1}\big(\Delta S_\text{BH}-S(D)\big)\ .
\label{fit}
}
At recovery, we will write $R_2=R^\natural\cup R^\sharp$; hence,
\EQ{
S_\text{rad}(R_2)\equiv S_\text{rad}(R^\natural\cup R^\sharp)=\frac\xi{\xi+1}\big(\Delta S_\text{BH}+S(D)\big)\ .
\label{fat}
}
One can view the time associated to the radiation in $R^\natural$, 
\EQ{
t\sim \frac{\Delta S_\text{BH}-S(D)}T\ ,
\label{pro}
} 
as the time it takes for the black hole to `process' the information in $D$ and start to emit its information in the radiation. The fact that there is a processing time was anticipated by Hayden and Preskill \cite{Hayden:2007cs} as the scrambling time of the black hole, an information theoretical measure of how long it takes for the information in $D$ to be widely distributed in the black hole state. This time scale, however, is of order $T^{-1}\log S_\text{BH}/c$ which in the present analysis is a subleading effect that we will analyse more carefully section \ref{s4} when we solve for the backreaction in JT gravity. However, we have found that the processing time also includes the contribution \eqref{pro} that involves the energy and entropy backreaction caused by $D$.

On the other hand, the time scale implicit in the radiation subset $R^\sharp$ is identified as the time it takes to emit radiation with an entropy $\sim S(D)$. Indeed, in the reversible limit $\xi\to1$, the entropy $S(R^\sharp)$ is precisely $S(D)$.

Notice that even though the radiation $R_1\cup R^\natural$ is not correlated with $\overline D$ it can be useful to decode $D$. This is because $R_1\cup R^\natural$ is entangled with the additional radiation $R^\sharp$. This corresponds to the fact that, in the Hayden-Preskill scenario \cite{Hayden:2007cs}, Bob collects all the radiation from when the black hole formed even though the early radiation never actually interacts with $D$. This is a characteristic feature of QECC, decoding is aided by having access to subsystems that never interacted with the encoded subsystem but which are entangled with subsystems that did.

\subsection{Recovery in the late radiation}
\label{sec:HP_late}

On the other hand, the fact that $I(R_1\cup R^\natural,\overline D)=0$ implies that $I((R_1\cup R^\natural)',\overline D)=2S(D)$, where the superscript prime indicates the complement in the set of all the Hawking radiation. In fact, we can be more specific by considering the conditions on an interval of late radiation $\tilde R=[u_1,u_2]$, with $u_1>t_0$, for which recovery is possible, $I(\tilde R,\overline D)=2S(D)$. Recovery will occur when $S(\tilde R)$ and $S(\tilde R\cup\overline D)$ are dominated by the $\tilde I=[t_0,u_2]$ saddle:
\begin{center}
\begin{tikzpicture} [scale=0.8]
\filldraw[fill = Plum!10!white, draw = Plum!10!white, rounded corners = 0.2cm] (-2.5,1.1) rectangle (6.6,-3.3);
%
\draw[decorate,very thick,black!40,decoration={snake,amplitude=0.03cm}] (0,-2.5) -- (0,0.5);
\draw[dotted,thick] (0,0) -- (6,0);
\draw[dotted,thick,red] (0,-1) -- (6,-1);
\draw[dotted,thick,blue] (0,-2) -- (6,-2);
\draw[dotted,thick,OliveGreen] (3,-2.5) -- (3,0.5);
\draw[dotted,thick] (4,-2.5) -- (4,0.5);
\draw[dotted,thick] (5,-2.5) -- (5,0.5);
\draw[very thick] (4,0) -- (5,0);
\filldraw[black] (5,0) circle (2pt);
\filldraw[black] (4,0) circle (2pt);
	\draw[very thick,red] (2.7,-1) -- (5,-1);
	\filldraw[red] (2.7,-1) circle (4pt);
	\filldraw[red] (5,-1) circle (4pt);
	\filldraw[OliveGreen] (3,-1) circle (2.5pt);
	\draw[OliveGreen] (3,-1) circle (4pt);
	\draw[very thick,blue] (3,-2) -- (4,-2);
	\filldraw[blue] (3,-2) circle (2pt);
	\filldraw[blue] (4,-2) circle (2pt);
	\node at (4,-2.8) {\footnotesize$u_1$};
	\node at (5,-2.8) {\footnotesize$u_2$};
	\node[right] at (-2.5,0) {$\tilde R$};
	\node[right] at (-2.5,-1) {$\tilde I$};
	\node[right] at (-2.5,-2) {$\tilde R\ominus\tilde I$};	
	\node at (4.5,0.4) {\footnotesize$\tilde R$};
	\node at (3.5,0.4) {\footnotesize$A$};
	\node at (3,-2.8) {\footnotesize $t_0$};
	\end{tikzpicture}
\end{center}
\noindent Once again this happen first for $S(R\cup\overline D)$ and then $S(R)$. During this interval, which can be written,
\EQ{
\frac\xi{\xi+1}\big(2S_\text{BH}(t_0)+\Delta S_\text{BH}&-S(D)\big)+\frac{\xi-1}{\xi+1}S_\text{rad}(A)\leq S_\text{rad}(\tilde R) \\ &\leq \frac\xi{\xi+1}\big(2S_\text{BH}(t_0)+\Delta S_\text{BH}+S(D)\big)+\frac{\xi-1}{\xi+1}S_\text{rad}(A)\ ,
}
where $A=[t_0+\epsilon,u_1]$, the mutual information rises until recovery is attained when $S_\text{rad}(\tilde R)$ equals the upper bound of the interval. It follows that recovery is always possible for a suitably small $A$ and large enough $\tilde R$ with an upper bound on the former
\EQ{
S_\text{rad}(A)<S_\text{rad}(R^\natural)\qquad\implies \quad A\subset R^\natural\ .
\label{pax}
}
This makes perfect sense because the subset $R^\natural$ is not correlated with $\overline D$ and so need not be included in $\tilde R$ in order to recover the information. Of course it can be included and does make it easier to recover $D$ because making $A$ smaller lowers the recovery time.

\subsection{Scrambling time}

Hitherto we have worked at leading order and ignored contributions to the entropy that are logarithmic in the Bekenstein-Hawking entropy. These corrections are interesting because, physically, they express the scrambling time of the black hole \cite{Hayden:2007cs}. The refinement of \eqref{twm} including the log corrections is derived in appendix \ref{app:A}. The corrections modify the contribution of the QES in \eqref{twm}:
\EQ{
S_\text{BH}(u_j)\longrightarrow S_\text{BH}(u_j)-\frac c{24}\log\frac{S_\text{BH}(u_j)}c\ .
\label{twm2}
}
The expression above is not valid for the QES that appears just before $D$ falls in. The correction for this QES  will be calculated in section \ref{s4}.

\subsection{A quantum error correcting code}

We have seen that the information in $D$ can be recovered both in $R=[0,u]$ and $\tilde R=[u_1,u_2]$. This kind of redundancy of information recovery is characteristic of a QECC. It is precisely why a QECC is robust against errors because one can obviously corrupt the complements $R'$ or $\tilde R'$ and still recover the information in $D$. Let us develop this connection further.

It is an important consistency condition that the subsets $R$ and $\tilde R$ have a non-trivial minimal overlap, in fact it is precisely the set $R^\sharp$,
\EQ{
R^\sharp\subset R\cap\tilde R\ .
}
At this point we could fall into a tempting fallacy and say that $R^\sharp$ must contain $D$'s information. But quantum information is subtle, as pointed out in the introduction, and actually the opposite is true: the information in $D$ cannot be recovered from $R^\sharp$ since $I(R^\sharp,\overline D)=0$: the subset $R^\sharp$ by itself is not big enough. Counter-intuitively, this implies that $D$'s information can actually be recovered in the complement $(R^\sharp)'$. This is just precisely the way that information is encoded in a simple QECC: out of the three subsets of the Hawking radiation $R_1\cup R^\natural$, $R^\sharp$ and $R'=B$, since the future radiation is equivalent to the remaining black hole $B$, whose union is the complete set of radiation, the information in $D$ can recovered in any pair of subsets but not in any single subset:
\begin{center}
\begin{tikzpicture} [scale=0.8]
\filldraw[fill = Plum!10!white, draw = Plum!10!white, rounded corners = 0.2cm] (-2.5,1.2) rectangle (7.6,-3.3);
\draw[decorate,very thick,black!40,decoration={snake,amplitude=0.03cm}] (7,-2.5) -- (7,0.5);
\draw[decorate,very thick,black!40,decoration={snake,amplitude=0.03cm}] (0,-2.5) -- (0,0.5);
\draw[dotted,thick] (0,0) -- (7,0);
\draw[dotted,thick] (0,-1) -- (7,-1);
\draw[dotted,thick] (0,-2) -- (7,-2);
\draw[dotted,thick,OliveGreen] (3,-2.5) -- (3,0.5);
\draw[dotted,thick] (4,-2.5) -- (4,0.5);
\draw[dotted,thick] (5,-2.5) -- (5,0.5);
\draw[very thick] (4,0) -- (0,0);
\filldraw[black] (0,0) circle (2pt);
\filldraw[black] (4,0) circle (2pt);
	\draw[very thick] (4,-1) -- (5,-1);
	\filldraw[black] (4,-1) circle (2pt);
	\filldraw[black] (5,-1) circle (2pt);
	\filldraw[OliveGreen] (3,-1) circle (2.5pt);
	\draw[OliveGreen] (3,-1) circle (4pt);
	\filldraw[OliveGreen] (3,-2) circle (2.5pt);
	\draw[OliveGreen] (3,-2) circle (4pt);
	\filldraw[OliveGreen] (3,0) circle (2.5pt);
	\draw[OliveGreen] (3,0) circle (4pt);	
	\draw[very thick] (5,-2) -- (7,-2);
		\filldraw[black] (5,-2) circle (2pt);
	\filldraw[black] (7,-2) circle (2pt);
	%
	%
	\node[right] at (-2.4,0) {$R_1\cup R^\natural$};
	\node[right] at (-2.4,-1) {$R^\sharp$};
	\node[right] at (-2.4,-2) {$R'=B$};	
	\node at (4.5,0.4) {\footnotesize$R^\sharp$};
	\node at (3.5,0.4) {\footnotesize$R^\natural$};
	\node at (2,0.4) {\footnotesize$R_1$};	
	\node at (0,-2.8) {\footnotesize $0$};
		\node at (3,-2.8) {\footnotesize $t_0$};
	\node at (7,-2.8) {\footnotesize $\infty$};
	\end{tikzpicture}
\end{center}
The structure here is exactly like the simplest QECC, namely the three qutrit QECC, reviewed in a holographic context in \cite{Harlow:2018fse}. In this case, we can think of one of the qutrits $Q_1$ as $D$ that interacts with $Q_2$, the black hole. $Q_3$ plays the r\^ole of the radiation previously emitted. The qutrit $Q_2$ is entangled with $Q_3$ but $Q_1$ never interacts with $Q_3$ (the radiation $Q_3$ has already dispersed). After time evolution generated by a unitary acting only on $Q_1$ and $Q_2$, the information contained in $Q_1$ can be recovered from any pair of qutrits but not from any single qutrit: $I(Q_j,\overline Q_1)=0$ while $I(Q_j\cup Q_k,\overline Q_1)=2S(Q_1)$. 

\section{Solving the backreaction in JT gravity}
 \label{s4}

For a more refined analysis, we need to able to solve for the backreaction of $D$ on the geometry as it falls in. In general, this is difficult problem and would need to be solved numerically. Remarkably, however, in the model of JT gravity the backreaction can be solved analytically. The analysis is particulary simple when $D$ is a localised packet of energy and momentum (i.e. a shockwave) which falls in along an ingoing null geodesic. The model is itself remarkable as it is simple enough to be tractable yet also captures the dynamics of the $s$-wave sector of the near-horizon limit of the near-extremal Reissner-Nordstr\"om (RN) black hole in $3+1$ dimensions \cite{Almheiri:2014cka,Nayak:2018qej}. What is particularly interesting for the present work is that the model can describe an evaporating black hole with arbitrary infalling matter. The focus will be on a localized packet of infalling matter $D$, carrying both energy and entropy.

\begin{figure}[ht]
\begin{center}
\begin{tikzpicture} [scale=1.2] 
\filldraw[pink!20] (4,4.4) -- (3.4,5) -- (-2,5) -- (-2,0) -- (-0.4,0) -- cycle;
\draw[purple!20,pattern=north east lines,pattern color=black,opacity=0.3] (3.5,1.5) -- (6,4) -- (7,3) -- (4.5,0.5) -- cycle;
\draw[dash dot,thick] (5,5) -- (5,0);
\draw[dotted,thick] (7.5,2.5) -- (5,0);
\node at (1.5,0.2) {\footnotesize AdS};
\node at (7.5,0.2) {\footnotesize Minkowksi};
\draw[very thick, dashed] (2,2) -- (5,5) node[sloped,midway,below] {\footnotesize $\tilde U=0$};
\draw[very thick, dashed] (-2,0) -- (2,4) node[sloped,midway,below] {\footnotesize $U=0$};
\filldraw[white] (10,0) -- (5,5) -- (10,5) -- cycle;
\draw[very thick] (10,0) -- (5,5);
\filldraw[green!10] (0.8,3.2) -- (-2,0.4) -- (-2,5) -- (-1,5) -- cycle;
\draw[decorate,very thick,decoration={snake,segment length=2mm,amplitude=0.4mm},dotted] (3.4,5) -- (-2,5);
\draw[decorate,very thick,decoration={snake,segment length=2mm,amplitude=0.4mm}] (5,5) -- (3.4,5);
\node[Emerald] at (-0.85,3.7) {$I_1$};
\node[red] at (1,4.5) {$I_2$};
\draw[blue,thick] (-2,3.2) to[out=-10,in=-170] (0.8,3.2);
\draw[blue,thick] (0.8,3.2) to[out=10,in=135] (6,4);
\filldraw[Emerald] (0.8,3.2) circle (0.15cm);
\filldraw[pink] (4,4.4) circle (0.15cm);
\draw[purple,line width=1mm,->] (5,0) -- (1.2,3.8);
\node[purple,rotate=-45] at (3,1.4) {$D$};
%
\node[rotate=-45] at (7.75,2.65) {\footnotesize $t_0$};
\draw[blue,line width=1mm] (10,0) -- (6,4);
\node[blue,rotate=-45] at (9.3,1.5) {$R_1$};
\node[blue,rotate=-45] at (7.6,3.8) {$R_2$};
\node[rotate=-45] at (8.8,3) {$\mathscr I^+$};
\node[Emerald,rotate=-45] (b1) at (-1,2.4) {$\overline R_1$};
\node[red,rotate=-45] at (1.5,2.6) {$\overline R_2$};
\draw[blue,thick,decorate,decoration = {brace}] (6.4,4.4) -- (7.9,2.9);
\node[blue,rotate=-45] at (7.45,2.95) {$R^\natural$};
\node[blue,rotate=-45] at (6.7,3.7) {$R^\sharp$};
\node[purple,rotate=45,right] at (4.3,1.4) {\footnotesize scrambled $D$ info};
\end{tikzpicture}
\caption{\footnotesize A suitable Cauchy slice (in blue) to calculate $S(\rho^\text{sc}_{R\cup I})$ the entropy of the semi-classical reduced state on $R\cup I$. The information recovery involves a competition of two saddles with QES and island in green $I_1$ and pink $I_2$. $R$ is split into $R_1\cup R_2$, the radiation submitted before $D$ crosses into the AdS region and $R_2$ afterwards. The island $I_1$ holds the purifier of $R_1$ and $I_2$ of $R_1\cup R_2$. When $I_1$ dominates $D$ experiences a smooth horizon before being scrambled and evaporated. After a hiatus $R^\natural\subset R_2$, the information in $D$ is recovered continuously during the interval $R^\sharp\subset R_2$.}
\label{fig3} 
\end{center}
\end{figure}
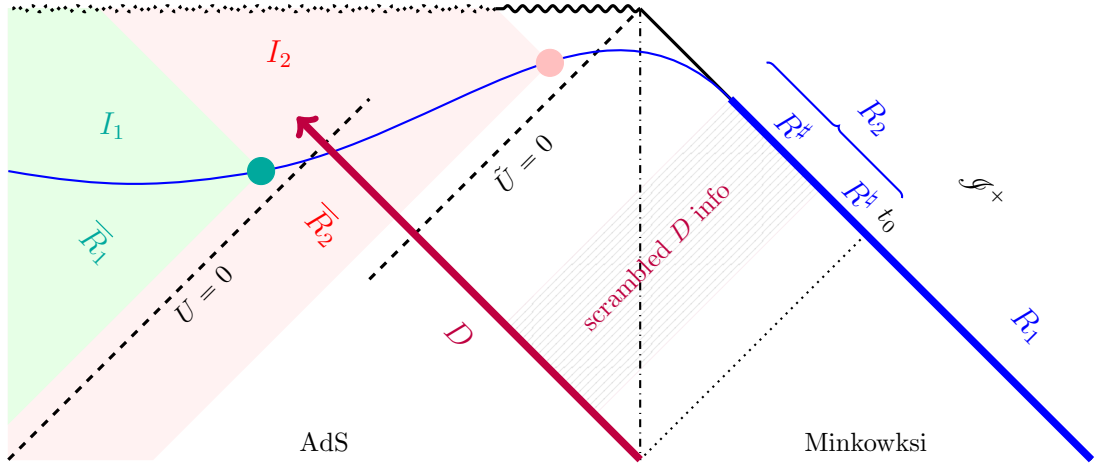

In the model, the geometry of the black hole is a dynamically determined patch of AdS$_2$ which is matched onto the boundary of a half Minkowski space, as described in \cite{Almheiri:2019psf}. The set up is shown in figure \ref{fig3}. The half Minkowski space region provides the auxiliary subsystem $R$ which collects the Hawking radiation. We can think of the radiation as being collected at $\mathscr{I}^+$ as would be the case for a black hole in asymptotically flat space. The matter is provided by a large-$c$ CFT, which, for simplicity, can be a theory of a free bosons or fermions. Large $c$ ensures that there is a semi-classical limit.

\subsection{Some aspects of black holes in JT gravity} \label{sec:someaspects}

Whilst the metric in JT gravity is fixed to be that of AdS$_2$, there is a dynamical scalar field $\phi$, the dilaton, which, in the semi-classical approximation, is sourced by the expectation value of the energy-momentum tensor. In the context of the RN black hole in $3+1$ dimensions, the dilaton is the area of the transverse $S^2$. It is convenient to express the dynamics of JT gravity in terms of the shape of the boundary curve \cite{Engelsoy:2016xyb,Maldacena:2016upp,Jensen:2016pah}. Up to isometry, the shape of the curve is specified by a single function $V(t)$, which relates the Minkowski null coordinates ($u=t-x, v=t+x$) of the bath region $x\geq0$ to the KS coordinates $(U,V)$ of the AdS$_2$ region, with
\EQ{
V=V(v)\ ,\qquad U=U(u)\equiv -1/V(u)\ . 
}
For later convenience, we will use the freedom to normalise $V(t)$ so that $V(t_0)=1$, which ensures that $D$ crosses the AdS-Minkowski interface at $V=-U=1$. Note that we use $V$ as a coordinate and also as the map $V(t)$ and for the latter we will write the argument explicitly. The metric in the AdS$_2$ and Minkowski regions are given, respectively, by
\EQ{
ds^2=-\frac{4dU\,dV}{(1+UV)^2}\ ,\qquad ds^2=-du\,dv\ .
\label{mcc}
}
The equation of motion for the dilaton can be solved exactly in terms of the map $V(t)$ in the case that $T_{UU}=0$,\footnote{Of course there will be a non-trivial flux of Hawking radiation $T_{uu}\neq0$ through the interface between AdS and Minkowski regions.} i.e. no outgoing matter at the horizon, but with arbitrary infalling matter:\footnote{In the following, $k=G_N\cc/3\phi_r$, where the constant $\phi_r$ determines the behaviour of the dilaton on the AdS/Minkowski interface.}
\EQ{
\phi(U,v)=\phi_0+\frac{2G_N\cc}{3k}\Big(\frac{V''(v)}{2V'(v)}-\frac{UV'(v)}{1+UV(v)}\Big)\ .
\label{lss}
}
Here, $\phi_0$ determines the extremal entropy $S_*=\phi_0/4G_N$ which, for simplicity, we are taking to be negligible. So, the solution boils down to finding $V(t)$ for which the steps are as follows:

\begin{enumerate}[label=\protect\circled{\arabic*}]
\item Solve for the ADM energy $E=M-M_*$ by matching the energy flow across the AdS-Minkowski interface\footnote{This is simply the equation of motion for the map $V(t)$ when $T_{UU}=0$.}
\EQ{
\frac{dE}{dt}=T_{vv}-kE\ ,
\label{ece}
}
where $T_{vv}$ is the normal-ordered ingoing components of the energy-momentum tensor, which can be interpreted as the energy flux of infalling matter, which, in this section, is taken to be a shockwave. The second term is the flux of the Hawking radiation across the boundary. In the above, $M_*$ is the mass of the extremal black hole.
\item The function $V(t)$ is then determined by the Schwarzian equation
\EQ{
\big\{V(t),t\big\}=-\frac{24\pi k}{\cc}E(t)\ ,
\label{sch}
}
whose general solution is given by
\EQ{
V(t)=\frac{A F_1(t)+B F_2(t)}{C F_1(t)+D F_2(t)}\ ,
\label{mcd}
}
with $AD-BC \neq 0$ and where $F_i(t)$ are two independent solutions of the linear second order ODE
\EQ{
F''-\frac{12\pi k}{\cc}E(t)F=0\ .
\label{soe}
}
\item The constants of integration $\{A,B,C,D\}$ are determined (up to overall scaling) by imposing the continuity of $V(t)$, up to its second derivative, across the shockwave of infalling matter. We assume that the black hole is formed by a shockwave of collapsing matter at $t=0$ and, similarly, $D$ is modelled as a shockwave sent in at a later time $t_0$.\footnote{Concretely, these shockwaves can be realised by a local quench on the boundary, as described in \cite{Hollowood:2020cou}.}
\end{enumerate}

In the case that there is no infalling matter after the black hole is formed $E=E_0e^{-k t}$ for $t>0$ and the solution of \eqref{soe} involves Bessel functions \cite{Almheiri:2019psf}
\EQ{
F_1(t)=K_0(z)\ ,\quad F_2(t)=I_0(z)\ ,\qquad z=\frac{2\pi T_0}ke^{-kt/2} \label{fullsolbessel}\ ,
}
where $T_0^2=\frac{\pi\cc}{12k}E_0$. There is a simpler approximate solution which is valid until very late times. More precisely, it is valid for time scales of order $k^{-1}$ but smaller than $k^{-1} \log(T_0/k)$. In this regime, which we refer to as the adiabatic regime, there is a WKB-like solution of \eqref{soe}
\EQ{
V(t)=\exp\Big(2\pi \int_{t_0}^t T(t)\,dt\Big)\ ,
\label{guy}
}
where 
\EQ{
T(t)=T_0e^{-kt/2}\ ,
}
can be interpreted as the instantaneous temperature of the Hawking radiation that passes through the boundary $x=0$ at time $t$. Note that, in this approximation, the energy depends on the temperature as
\begin{equation}
    E(t)= \frac{\pi c}{12 k} T(t)^2\ .
\end{equation}
The Bekenstein-Hawking entropy is given by evaluating the dilaton on the horizon, which is located at $U=-1/V(\infty)=0$,
\EQ{
S_\text{BH}(t)=\frac{\phi(U=0,v=t)}{4G_N}=S_*+\frac\cc{6k}\frac{V''(t)}{2V'(t)}\approx S_*+\frac{\pi\cc}{12}\int_t^{t_\text{evap}} T(t')\,dt'\ , 
\label{pp1}
}
where the last expression is valid in the adiabatic limit. Alternatively, \eqref{pp1} can be derived by integrating the thermodynamic relation $dE=T\,dS$. In the above, $t_\text{evap}$ is the evaporation time which is $\infty$ in JT gravity and for the near-extremal RN black hole, while it is finite in the Schwarzschild case. Note that in JT gravity the black hole entropy is linear in $T$, as expected for a near-extremal black hole. Indeed, integrating \eqref{pp1} we find
\begin{equation}
    S_\text{BH}(t) = S_*+ \frac{\pi c}{6 k} T(t) \, .
    \label{eq:S_BH_JT}
\end{equation}
From this equation it is clear that the adiabatic approximation can be rephrased in more general terms as:
\EQ{
S_\text{BH}(t)\gg\cc\ .
\label{oss}
}
In our analysis, we will assume that the extremal entropy is small compared with other entropies in the problem so that we can neglect its contribution, effectively sending $S_*\to0$. Notice that this is not the limit that is related to the near-extremal RN black hole in $3+1$, but it simplifies the analysis. In this limit, we can introduce an analogue of a space-like singularity which is signalled by the vanishing of the dilaton, i.e.~the curve
\EQ{
U=\big(2V'(v)^2/V''(v)-V(v)\big)^{-1}\ ,
}
which is approximately $UV\approx 1$ in the adiabatic regime.

Before concluding this review, we consider the entropy of a single interval in the bath defined by a spacelike surface with endpoints $(u_1,v_1),(u_2,v_2)$, $u_2>u_1$. The regularised entropy, when there is no island, is given by the formula \cite{Calabrese:2004eu}
\EQ{
S(\rho_{R})=\frac\cc6\log\frac{(U_1-U_2)(v_2-v_1)}{\Omega_1\Omega_2}\ ,
}
where $\Omega_i = \sqrt{U'(u_i)}$ are the conformal factors that result from writing the metric in the Unruh state i.e. the vacuum in the mixed frame, $ds^2=-\Omega^{-2}dU\,dv$. We consider the contribution from the infalling vacuum modes $\sim c\log(v_2-v_1)$ to be subleading so that it can be safely ignored. In other words, to leading order, it does not matter what we take for the infalling coordinate of the endpoints of $R$ and so we choose to show $R$ close to $\mathscr I^+$. Ignoring the contribution of the infalling modes, we have
\EQ{
S(\rho_R) = \frac{c}{6} \log \frac{\sinh \pi \int_{u_1}^{u_2} T(t) dt}{\pi \sqrt{T(u_1)T(u_2)}} \approx \frac{\pi c}{6} \int_{u_1}^{u_2} T(t) dt = S_\text{rad}(R)\ ,
\label{pp2}
}
where the approximation holds when the argument of $\sinh$ is large, which corresponds to $u_2-u_1 \gg O(T^{-1})$. From now on, we will automatically neglect terms that are small in the adiabatic approximation, replacing $\approx$ with an equality.
Equation \eqref{pp2} ignores the effect of a greybody factor which in this model is a consequence of choosing transparent boundary conditions for the matter fields across the AdS-Minkowski interface. In this section we set the greybody coefficient (introduced in \eqref{eq:irr}) $\xi=2$. It is possible to relax this condition at the expense of a more complicated analysis: see \cite{Hollowood:2021lsw}.

\subsection{Including an infalling system}

As $D$ falls in, during an interval of null time $v \in[t_0,t_0+\epsilon]$ with $\epsilon \to 0$, the energy of the black hole jumps discontinuously according to $E(t_0+\epsilon)=E(t_0)+\cal{E}_D$. Consequently, the temperature also jumps
\EQ{
\frac{\pi c}{12 k} T(t_0+\epsilon)^2 = \frac{\pi c}{12 k} T(t_0)^2 + \cal{E}_D\ .
\label{eq:energy_after_diary}
}
Similarly, the map $V(t)$ is no longer smooth across $t=t_0$ but only continuous up to its second derivative. Using the recipe described in section \ref{sec:someaspects}, the new solution is related to the adiabatic one \eqref{guy} by a M\"obius transformation
\EQ{
V(t)=\frac{\lambda+\exp\Big(2\pi\int_{t_0}^tT(t)\,dt\Big)}{1+\lambda\exp\Big(2\pi\int_{t_0}^tT(t)\,dt\Big)}\ ,\qquad t>t_0\ ,
\label{hee}
}
where
\EQ{
\lambda=\frac{T(t_0+\epsilon)-T(t_0)}{T(t_0+\epsilon)+T(t_0)}\ .
}
The change in the map $V(t)$ induced by $D$ in \eqref{hee} suggests that we should introduce a new set of KS-type coordinates $(\tilde U,\tilde V)$ with an associated map
\EQ{
\tilde V(t)=\exp\Big(2\pi\int_{t_0}^t T(t)\,dt\Big)\ ,
}
related to $(U,V)$ by the M\"obius transformations 
\EQ{
\tilde V=\frac{V-\lambda}{1-\lambda V}\ ,\qquad \tilde U=\frac{U+\lambda}{1+\lambda U}\ .
}
Since Möbius transformations are isometries of AdS$_2$, the metric takes the same form as \eqref{mcc} when written in terms of these new coordinates.

The significance of these new coordinates is that they are adapted to the true event horizon of the black hole $\tilde U=0$ i.e.~$U=-\lambda$, and not the original horizon $U=0$ without shockwave. One is tempted to say that $D$ causes the horizon to jump out as it is absorbed by the black hole. However, this is not correct as the event horizon is a teleological concept so was never really at $U=0$.\footnote{However, the apparent horizon moves after the shockwave.} Note that for the original map $V(t)$, $D$ causes the asymptotic behaviour to change to $V(\infty)=1/\lambda$, i.e.~$\tilde V=\infty$.
The shift in the black hole entropy with or without the shockwave, introduced in section \ref{sec:IITS_with_D}, can be explicitly computed using \eqref{eq:S_BH_JT} and \eqref{eq:energy_after_diary}:
\EQ{
\Delta S_\text{BH} = \frac{\phi(\tilde{U}\!=0,v\!=\!t_0\!+\!\epsilon)}{4G_N}-\frac{\phi(U\!=\!0,v\!=\!t_0\!)}{4G_N} = S_\text{BH}(t_0) \left(\frac{T(t_0+\epsilon)}{T(t_0)}-1\right) \approx \frac{\mathcal{E}_D}{T(t_0)}\ ,
\label{eq:JT_1law_der}
}
which is consistent with the first law \eqref{eq:1st_law} in the limit where the diary energy is much smaller than initial energy of the black hole. In the same approximation, we can use \eqref{eq:JT_1law_der} to rewrite $\lambda$ in terms of the initial black hole entropy and $\Delta S_\text{BH}$:
\EQ{
\lambda = \frac{\Delta S_\text{BH}}{2S_\text{BH} + \Delta S_\text{BH}}\approx \frac{\Delta S_\text{BH}}{2S_\text{BH}(t_0)}\ll 1 \, .
\label{yer}
}

\subsection{Islands}

Consider the entropy of the Hawking radiation collected at $\mathscr I^+$ in the interval of null time $R=[-\infty,u]$. We assume that the black hole was formed by a shockwave sent into the extremal black hole at $t=0$. As shown in \cite[sec. 6]{Hollowood:2020cou}, for early times, the entropy of $R$ is given by
\EQ{
S(R)=S_{I_0}(R)= S_*+S_\text{rad}(R)\approx S_\text{rad}(0,u)\ ,
}
where $I_0$ is an island that stretches from a QES in front of the shockwave which forms the black hole (i.e. with ingoing coordinate $v<0$) to join up with $R$ at spatial infinity. As we are neglecting the contribution of the extremal entropy we see that the $S_{I_0}(R)$ is simply given by the thermal entropy of the radiation that crosses the AdS-Minkowski interface from the formation of the black hole at $t=0$ up to time $t=u$.

At later times, it becomes favourable to have an island with a QES behind the shockwave i.e. with ingoing coordinate $v>0$. To investigate this possibility, we have to extremize the generalized entropy \eqref{hys}. First of all, the exact expression for the dilaton is given in \eqref{lss}. This gives the area term in \eqref{hys}. The term $S(\rho^\text{sc}_{R\cup I})$ is the entropy of the Unruh state on the interval $R\cup I$. A suitable Cauchy slice is shown in figure \ref{fig3}. This entropy can again be computed using the formula
\EQ{
S(\rho^\text{sc}_{R\cup I})=\frac\cc6\log\frac{(U_1-U_2)(v_2-v_1)}{\Omega_1\Omega_2}\ ,
}
where $(U_1,v_1)$ are the coordinates of the QES and $(U_2,v_2)$ of the endpoint of $R$. The conformal factors $\Omega_i$ for the AdS$_2$ and flat metric are
\EQ{
\Omega_1^{-2}=\frac{4V'(v_1)}{(1+U_1V_1)^2}\ ,\qquad \Omega_2^{-2}=\frac1{U'(u_2)}\ .
}
We now re-label the coordinates of the QES $(U_1,v_1)\to(U,v)$ and the coordinate for the endpoint of $R$, $U_2\to U(u)$. Hence, up to a constant and a UV divergence, we have 
\begin{equation}
    S_\text{gen}(U,v)= \frac{\cc}{6k}\left\{\frac{V''(v)}{2V'(v)}-\frac{UV'(v)}{1+UV(v)}\right\}+\frac\cc6\log\sqrt{\frac{V'(v)}{U'(u)}}\cdot\frac{U-U(u)}{1+UV(v)}+{\mathbb S}_D(v)\ .
\label{uee}
\end{equation}
We have also dropped the contribution to the entropy from the ingoing modes as they turn out not to be important in the extremization. The final term accounts for the entropy of $D$ and is given by
\EQ{
{\mathbb S}_D(v)=\begin{cases}\int_{0}^v{\mathfrak s}(t)\,dt \ ,& R\ ,\\ S(D)-\int_{0}^v{\mathfrak s}(t)\,dt\ , & R\cup\overline D\ ,\end{cases}
\label{iuu}
}
where ${\mathfrak s}(t)$ is the entropy flux of the diary, which is supported in the interval $[t_0,t_0+\epsilon]$, and the two cases correspond to whether the purifier $\bar D$ of $D$ is included or not. This term accounts for the overlap of $D$ with the island and adds to the entropy for $R$ and subtracts from the entropy for $R\cup\overline D$. For a shockwave, the entropy flux is given by
\EQ{
\mathfrak{s}(t) =S(D) \delta(t-t_0)\ ,
}
so that, for $R$, ${\mathbb S}_D(v)$ is $S(D)$ if the shockwave passes through the island and $0$ otherwise. For $R \cup \bar D$, the opposite is true.

For $u, v<t_0$, the backreaction of $D$ is irrelevant and the generalized entropy simplifies to
\EQ{
S_\text{gen}(U,v)\approx S_\text{BH}(v)(1-2UV(v))+\frac\cc6\log\sqrt{\frac{V'(v)}{U'(u)}}(U-U(u))+{\mathbb S}_D\ .
\label{eq:S_gen3}
}
This form assumes that the QES lies close to the horizon, in the sense that $UV\ll1$, which is established ex-post facto.\footnote{Notice that equation \eqref{eq:S_gen3} is written in such a way that it will apply to the $s$-wave sector of any black hole including the Schwarzschild black hole.}
Extremizing with respect to $U$ and $v$ determines the position of the QES
\EQ{
U=-\frac{U(u)}3\ ,\qquad V(v)=-\frac{\cc}{16S_\text{BH}(v)}\cdot\frac1{U(u)}\ .
\label{pcv}
}
From this, the expression for $UV$ quoted earlier \eqref{yet} follows. Since the black hole is evaporating slowly, we have the approximate solution \eqref{sum} and since $UV\ll1$, the QES lies just inside the horizon.

We now turn to the entropy at the extremum. In the first instance, in addition to the leading order contributions that are of order $S_\text{BH}$, we will keep log terms that are of order $\cc\log(S_\text{BH}/\cc)$ and $\cc\log(S_\text{BH}/\Delta S_\text{BH})$. These are associated to time scales of order the scrambling time of the black hole. Note that the log terms and scrambling times involve the entropy and temperature and these can be evaluated either at $t_0$ or $u$ since the difference is beyond the order to which we are working, therefore, we will not specify the arguments of these terms. Other terms beyond these are subleading.

Keeping only the terms described above, we have
\EQ{
&S_\text{BH}(v)\approx S_\text{BH}(u)+\frac\cc{24}\log\frac{S_\text{BH}(u)}\cc\ ,\\
&\frac\cc6\log\sqrt{\frac{V'(v)}{U'(u)}}(U-U(u))\approx -\frac\cc{12}\log\frac{S_\text{BH}(u)}\cc\ ,
}
and so the entropy of this saddle is 
\EQ{
S_{I_1}(R) = S_\text{BH}(u)-\frac\cc{24}\log\frac{S_\text{BH}(u)}\cc\ .
\label{vvz}
}
To adapt this expression for $R\cup\overline D$, simply add $S(D)$.

Now suppose that $u$ increases beyond $t_0$, the time on the boundary where $D$ falls in. The solution \eqref{pcv} remains valid as long as $v<t_0$. However, we can no longer use the approximation \eqref{guy} as the map $U(u)=-1/V(u)$ is modified when $u>t_0$ as in \eqref{hee}. 


It will important for the following analysis that for $u>t_0$, i.e.~$|\tilde U(u)|<1$, and taking account that $\lambda\ll1$, we have
\EQ{
U(u)\approx\tilde U(u)-\lambda
\label{oed}
}
and so as $u$ increases from $t_0$, for time scales that are large compared with the thermal time scale, the log of minus the right-hand side is effectively a sharp crossover 
\EQ{
\log(\lambda-\tilde U(u))\approx -\min\Big(\log\frac{S_\text{BH}(t_0)}{\Delta S_\text{BH}},2\pi\int_{t_0}^uT(t)\,dt\Big)\ .
\label{llk}
}
It then follows that the $v$ coordinate of the QES is approximately equal to
\EQ{
v=\min\Big(u-\frac1{2\pi T(u)}\log\frac{S_\text{BH}(u)}\cc,t_0-\frac1{2\pi T(t_0)}\log\frac{\Delta S_\text{BH}}\cc\Big)\ .
}
So as $u$ increases, the $v$ coordinate of the QES eventually freezes in front of $D$, $v<t_0$, with coordinates
\EQ{
v\to t_0-\frac1{2\pi T(t_0)}\log\frac{\Delta S_\text{BH}}\cc\ ,\qquad U\to \frac{\Delta S_\text{BH}}{6S_\text{BH}(t_0)}\ ,
}
which is the same result obtained in \cite[eq. (92)]{Penington:2019npb}. Notice that the inequality \eqref{rvv} is necessary for having $v<t_0$, otherwise we would have that the QES never freezes before the shockwave; for smaller diaries we would have that the shockwave backreaction, in our approximations, is negligible.

Using this and splitting the radiation $R=R_1\cup R_2$, where $R_1=[-\infty,t_0]$ and $R_2=[t_0,u]$, we have\footnote{In deriving this expression, we used $\tilde U'(u)\approx 2\pi T\tilde U(u)$ and $S_\text{rad}(R_2)=-c/12\log \tilde U(u)$.}
\EQ{
S_{I_1}(R)= \max\Big(S_\text{BH}(u)&-\frac\cc{24}\log\frac{S_\text{BH}(u)}\cc\,,\\[5pt]&S_\text{BH}(t_0)+\frac{\cc}{8}\log\frac{\Delta S_\text{BH}}\cc-\frac\cc{6}\log\frac{S_\text{BH}(t_0)}\cc+S_\text{rad}(R_2)\Big)\ .
}
The second term here wins out eventually and then the QES freezes as above. For $R\cup\overline D$ one simply adds $S(D)$ since $D$ does not lie in $I_1$.

When $u>t_0$ there is a new saddle with a QES that is behind $D$, i.e.~with $v>t_0$. This is simply the solution \eqref{pcv} in terms of the coordinates $(\tilde U,\tilde V)$. It is important that the generalized entropy takes the same form in terms of these coordinates because they are related to the old coordinates by a M\"obius transformation. The new saddle requires that $u>t_0+(2\pi T(t_0))^{-1}\log S_\text{BH}(t_0)/\cc$ in order that $v>t_0$. 
The entropy is as in \eqref{vvz} but now $D$ lies in the island and so
\EQ{
S_{I_2}(R)= S_\text{BH}(u)-\frac\cc{24}\log\frac{S_\text{BH}(u)}\cc+S(D)\ .
\label{wwz}
}
For the case $R\cup\overline D$, one removes the term $S(D)$ since now the island contains $D$, the purifier of $\overline D$.

\subsection{Information recovery}

Now we can discuss information recovery by evaluating the mutual information $I(R,\overline D)$. We assume that the black hole has evaporated past the Page time so that the island $I_1$ saddle dominates for $u<t_0$. In this case, it is immediately apparent that $I(R,\overline D)=0$. As $u$ increases through $t_0$ the correlation continues to vanish until the island $I_2$ is favoured over $I_1$ for $S(R\cup\overline D)$. The upper bound for this vanishing correlation defines $R_2=R^\natural=[t_0,u]$ by 
\EQ{
S_\text{rad}(R^\natural)=\frac23\big(\Delta S_\text{BH}-S(D)\big)+\frac{\cc}{12}\log\frac{S_\text{BH}}{\Delta S_\text{BH}}\ .
\label{rhd2}
}
This is a refinement of \eqref{fit} to include the log corrections for the case $\xi=2$. We have supressed the argument from the $\log S_\text{BH}$ term since the difference between $t_0$ and $u$ is very small because the black hole evaporates very slowly in the adiabatic regime.

The time interval associated to $R^\natural$ is interpreted as the time delay for the black hole to `process' $D$ and start to return its information via the radiation. The refinement above has contributions that are logarithmic in the black hole's entropy as anticipated by Hayden and Preskill \cite{Hayden:2007cs} but there is also a backreaction effect that we noted already in the adiabatic analysis of section \ref{s2}. 
As $u$ increases further, the mutual information rises until $R_2=R^\natural\cup R^\sharp=[t_0,u]$. Then, the saddle for $S(R)$ transitions $I_1\to I_2$ and, hence, $I(R,\overline D)=2S(D)$, with
\EQ{
S_\text{rad}(R^\natural\cup R^\sharp)=\frac23\big(\Delta S_\text{BH}+S(D)\big)+\frac{\cc}{12}\log\frac{S_\text{BH}}{\Delta S_\text{BH}}\ ,
\label{rhd}
}
which is a refinement of \eqref{fat}.

\section{Python's lunch}\label{s5}

In the analysis so far, we have not needed to resolve what happens to the generalized entropy when the would-be QES is in the $D$ interval $v\in[t_0,t_0+\epsilon]$ because the QES that dominate the entropy do not lie in this interval. However, it is also important to investigate QES that are maxima of the generalized entropy, and these will lie in the $D$ interval. These maxima are related to the complexity of the decoding the information in the emitted radiation, and have an important interpretation as `python's lunches' \cite{Brown:2019rox,Engelhardt:2021qjs}. These are configurations of two minimal QES with a maximum QES in between along a Cauchy slice behind the horizon. 

The intuition comes from thinking of the black hole geometry, or more precisely the Einstein-Rosen bridge, as a tensor network. The notion of decoding the state of the radiation can be viewed as the process of shortening the tensor network by acting with unitaries and also performing post selections. It is the latter that act as an obstruction to decoding and determine the exponential complexity.
With this interpretation, the height of the lunch is conjectured to quantify the complexity of decoding information that has been evaporated out of the black hole. 

Specifically, in the case that we are interested in, information allowing the recovery of $D$ is scrambled and evaporated out when the minimal QES exchange their dominance. After this time, $I(R,\overline D)=2S(D)$ is maximal and the information in $D$ is available to be decoded in $R$. This means that there is some decoding unitary $U_\text{dec}$ acting on $R$ that distills the entanglement with the purifier $\overline D$ into some convenient subspace $\widetilde D\subset R$. 
Finding a suitable $U_\text{dec}$ is expected to be a complex operation and the python's lunch configuration of the QES quantifies this complexity. The conjecture of \cite{Brown:2019rox} is that the complexity of decoding is dominated by the exponential behaviour
\EQ{
{\cal C}(U_\text{dec})\thicksim \exp\big[\big(S_\text{max}-S_\text{min}\big)/2\big]\ ,
\label{hyd}
}
where we are neglecting the subleading proportionality factor. In the above, $S_\text{max}$ is the generalized entropy for $R$ at the maximal QES that lies between the two minima. The entropy $S_\text{min}$ is defined as the minimum saddle with larger entropy. 

In this section, we want to quantify the complexity of decoding the diary alone, assuming that we have already decoded all the previously emitted radiation, or, equivalently, that we have shortened the Einstein-Rosen bridge up to the moment the diary falls inside the black hole. Since this lunch is given by a local maximum, we suggest to call it the `python's snack', to differentiate it from the main one which is a global maximum.
The interesting case is when we take $t_0$ after the Page time.
In order to decode the diary we need $R$ to be past the recovery time, therefore we have that the QES stuck before the diary has higher entropy, which means that in \eqref{hyd} $S_\text{min}=S_{I_1}(R)$. In the next section we determine, both analytically in the shockwave limit and numerically for finite $\epsilon$, the location of the maximal QES and $S_\text{max}$.

\subsection{Finding the maximum QES}

The infall of $D$ in the $\epsilon\to0$ limit creates a discontinuity in the derivative of the generalized entropy, due to the fact that the function $V(t)$ is only continuous up to the second derivative and the dilaton contains a term proportional to $V''(v)$, see equation \eqref{lss}. This means that, whilst $\partial_US_\text{gen}$ is continuous, $\partial_vS_\text{gen}$ is discontinuous across $v=t_0$. We will now argue that this behaviour means there should be another QES which is a maximum of the generalized entropy. We will then investigate the existence of the QES maximum numerically after smoothing out $D$.

We first solve $\partial_U S_\text{gen} = 0$ for $U$ and substitute the result into $\partial_v S_\text{gen}$.
For both the QES before and after the shockwave, we find
\begin{equation}
- \frac{12}{c} S_\text{BH}(t_0)+ \frac{1}{U-U(u)}=0 \, , 
\end{equation} 
where we have used $V = \tilde{V} = 1$. From this we get
\begin{equation}
    U= \frac{c}{12 S_\text{BH}}+U(u) \, .
\label{eq:U:QES_max_sol}
\end{equation}
We now use this result in the extremization of $S_\text{gen}$ with respect to $v$ for the QES before the diary
\begin{equation}
\partial_{v} S_\text{gen} = 2 \pi T(t_0) \left(\frac{c}{24}-2 S_\text{BH}(t_0) U V\right) = 2 \pi T(t_0) \left(2 S_\text{BH}(t_0) |U(u)|-\frac{c}{8}\right) \label{eq:dS_python11}
\end{equation}
and after it
\begin{equation}
\partial_{v} S_\text{gen} =  2 \pi T(t_0+\epsilon) \left(\frac{c}{24}-2 S_\text{BH}(t_0 + \epsilon) \tilde{U} \tilde{V}\right) =  2 \pi T(t_0+\epsilon) \left(2 S_\text{BH}(t_0) (|U(u)|-\lambda)-\frac{c}{8}\right)\, . \label{eq:dS_python22}
\end{equation}
Now we see that in the interval
\begin{equation}
\frac{c}{16 S_\text{BH}(t_0)}<|U(u)|<\frac{c}{16 S_\text{BH}(t_0)} + \lambda ,
\end{equation}
\eqref{eq:dS_python11} is positive while \eqref{eq:dS_python22} is negative. This indicates the presence of a maximum, i.e., a python's lunch. Notice that, in terms of the $\tilde{U}(u)$ coordinates, the lunch starts a scrambling time after the shockwave $u = t_0 + \Delta t_{\text{scr}.}$ and it is present for every time after it if $|U(u)|\ge\lambda > \frac{c}{16 S_{\text{BH}}}$, i.e., $\Delta S_\text{BH}> c/8$\footnote{Notice that this is again consistent with \eqref{rvv}.}.

In order to be more precise about the maximum of the python's lunch, we perform a numerical analysis of the exact generalized entropy \eqref{uee} with $D$ modelled as either a pulse with constant energy and entropy density in the interval $[t_0,t_0+\epsilon]$ or as an operator quench in the matter sector. For the former, the solution in this interval is now built out of the Bessel functions $Y_\nu(\nu e^{-kt/2})$ and $J_\nu(\nu e^{-kt/2})$ where
\EQ{
\nu=\frac{4}k\sqrt{\frac{3\pi{\cal E}_D}{\cc\epsilon}}\ .
}
At $t=t_0$ and $t=t_0+\epsilon$ one matches the solution to the adiabatic solutions with an arbitrary M\"obius transformation in order to ensure the continuity of $V(t)$ up to the second derivative. The Schwarzian equation \eqref{sch} then ensures that the third derivative is continuous. This will be sufficient to ensure that the generalized entropy is once differentiable.

When $D$ is modelled as an operator quench, as in \cite{Hollowood:2020cou}, the energy density has profile
\EQ{
T_{vv}(v)= \frac{2h}{\pi\epsilon}  \cdot\frac{\epsilon^3}{((v-t_0)^2 + \epsilon^2)^2}\ ,
}
where $\epsilon$ is a regulator and $h$ is the conformal dimension of the operator.

In both models, the map  $V(v)$ is found numerically. It is important that extremizing with respect to $U$ can be done exactly giving the $U$ coordinate of the QES
\EQ{
U=-\frac{k+k U(u)V(v)+U(u)V'(v)}{k V(v)+k U(u)V(v)^2-V'(v)}\ .
\label{umm}
}
So, the $U$ coordinate of the QES is determined exactly and then the generalized entropy can be plotted
off shell as a function of $v$ for various values of $u$, the coordinate of the endpoint of $R$ on $\mathscr I^+$. For some indicative values of the underlying parameters these off-shell plots are shown in figure \ref{fig4} for the simple energy pulse. In both cases, the minimal QES that exists in front of $D$ for $u<t_0$ is clearly visible and, as $u$ increases beyond $t_0$, a second minimal QES appears behind $D$. At a later time this has the lowest entropy. The exchange of minima indicates that the information allowing the recovery of $D$ has been evaporated out of the black hole. It is then obvious that there must be a maximum QES in between and that this lies in, or on the boundary of, the $D$ interval. 

\pgfdeclareimage[interpolate=true,width=7.5cm]{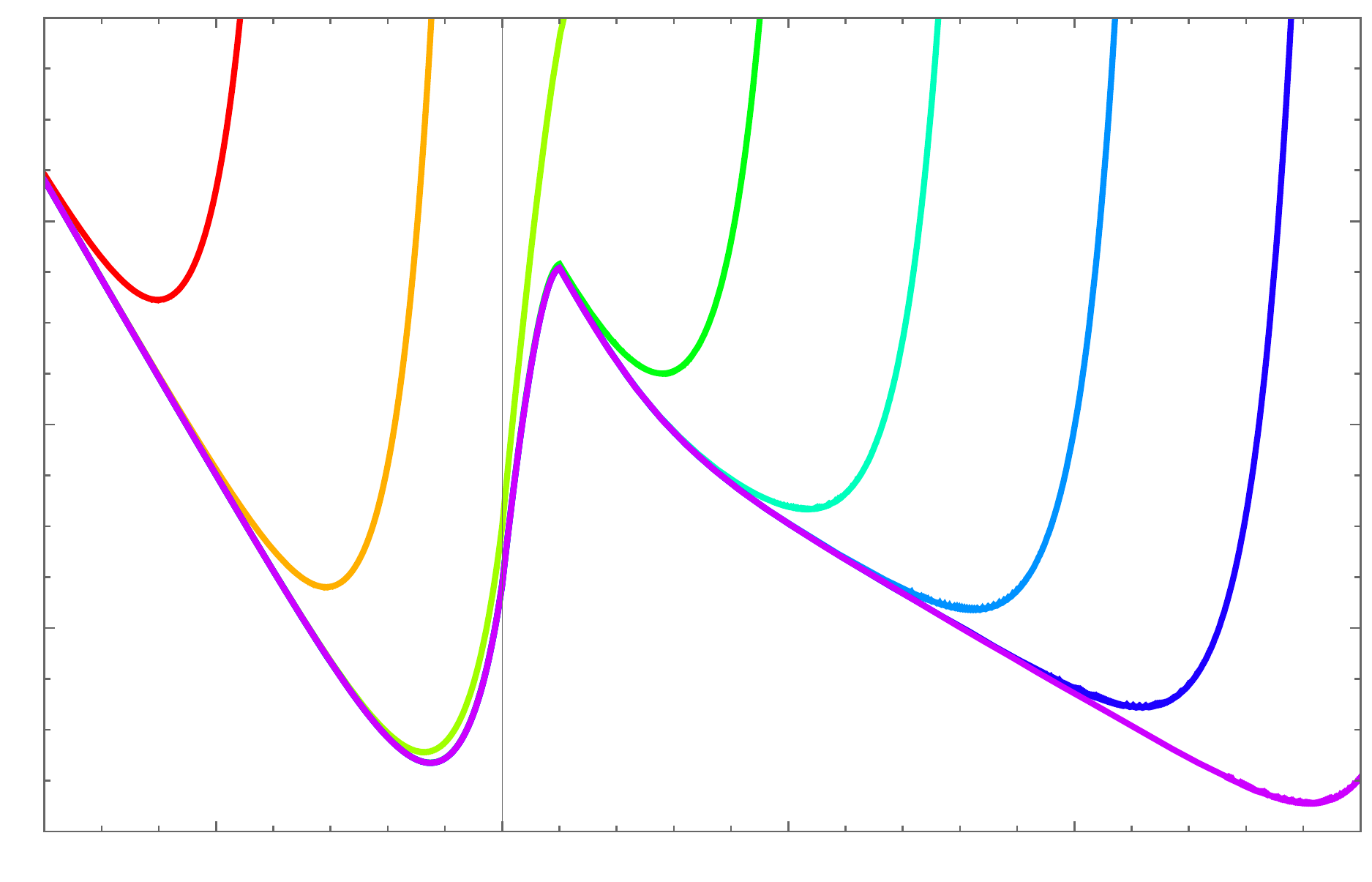}{pic1}
\pgfdeclareimage[interpolate=true,width=7.5cm]{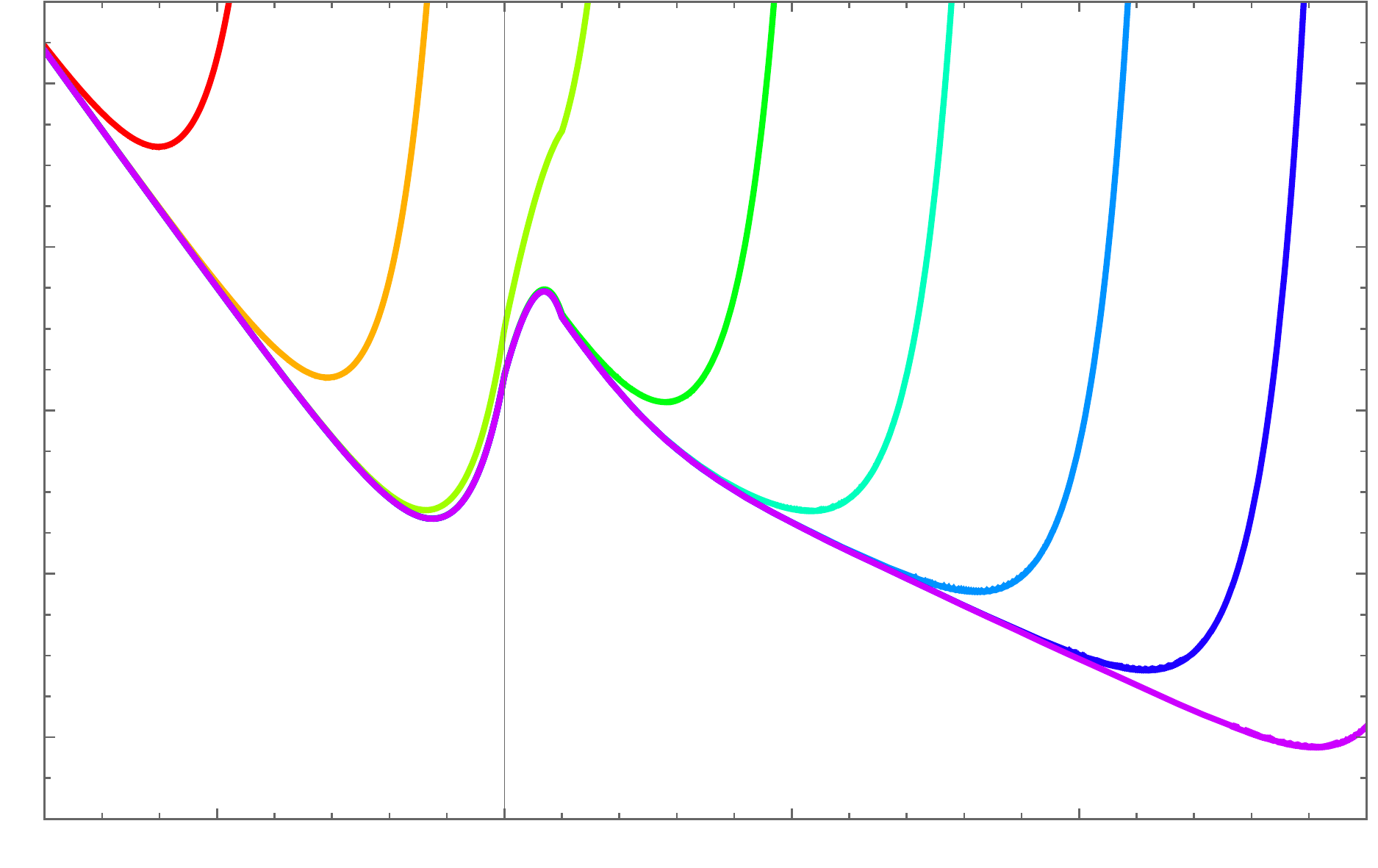}{pic2}
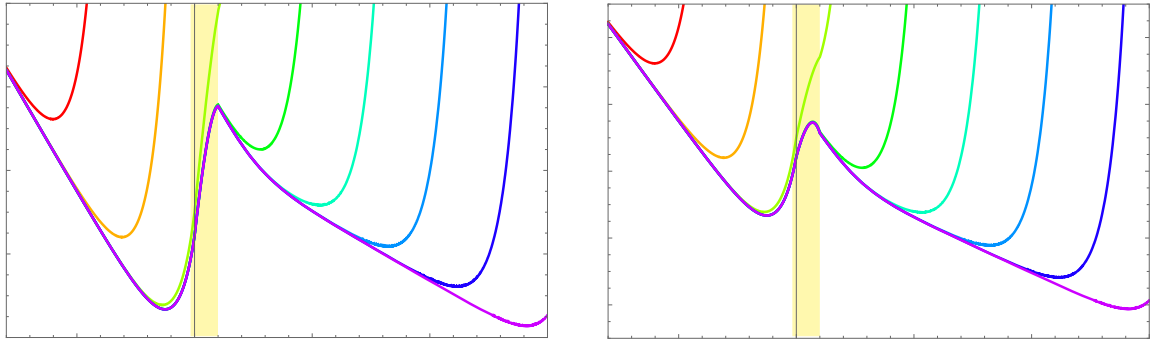
\begin{figure}
\begin{center}
\begin{tikzpicture}[scale=1]
\draw[yellow!40,fill=yellow!40] (2.7,0.3) rectangle (3.05,4.7);
\begin{scope}[xshift=8cm]
\draw[yellow!40,fill=yellow!40] (2.7,0.3) rectangle (3.05,4.7);
\end{scope}
\pgftext[at=\pgfpoint{0cm}{0cm},left,base]{\pgfuseimage{pic1}} 
\pgftext[at=\pgfpoint{8cm}{0cm},left,base]{\pgfuseimage{pic2}} 
\end{tikzpicture}
\caption{\footnotesize The off-shell generalized entropy for $R$ (left) and $R\cup\overline D$ (right) for the square energy pulse model as a function of $v$ (with the $U$ coordinate on-shell) for different values of the $u$ coordinate of the endpoint of $R$ on $\mathscr I^+$ indicated by the colour from red through purple. The $D$ interval of energy density is the shown as the yellow region. When $u<t_0$ (red) there is a QES in front of $D$, $v<t_0$. As $u$ increase beyond $t_0$ another QES develops with $v>t_0$. At a later $u$ this QES has the minimum entropy. It is also clear that there is a maximum QES with a coordinate $v$ inside the $D$ interval. The configuration of 3 QES is a python's lunch.}
\label{fig4}
\end{center}
\end{figure}

For the operator quench the results are very similar, see figure \ref{fig5}, showing that the behaviour is rather universal. 
\pgfdeclareimage[interpolate=true,width=7.3cm]{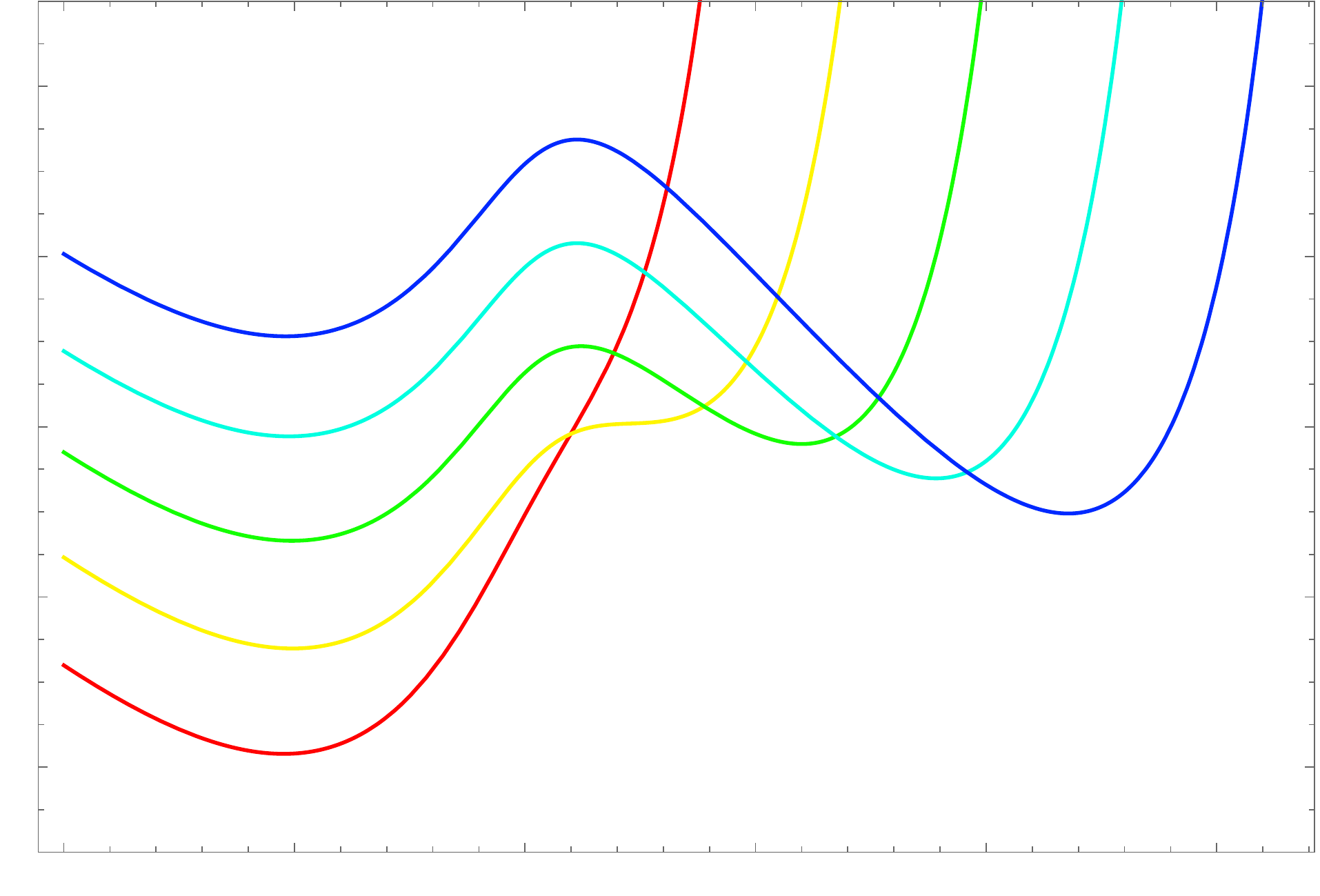}{pic3}
\pgfdeclareimage[interpolate=true,width=7.3cm]{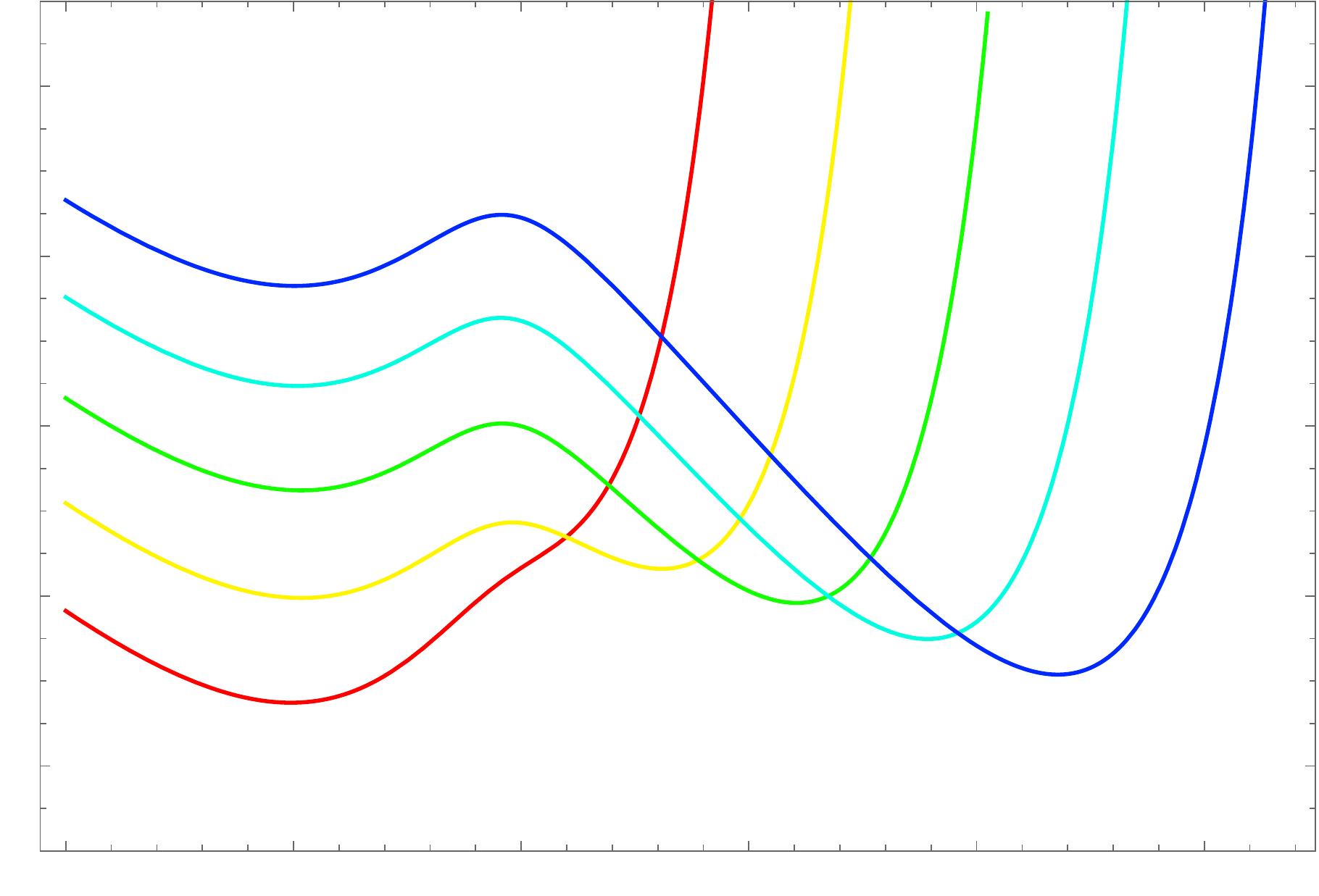}{pic4}
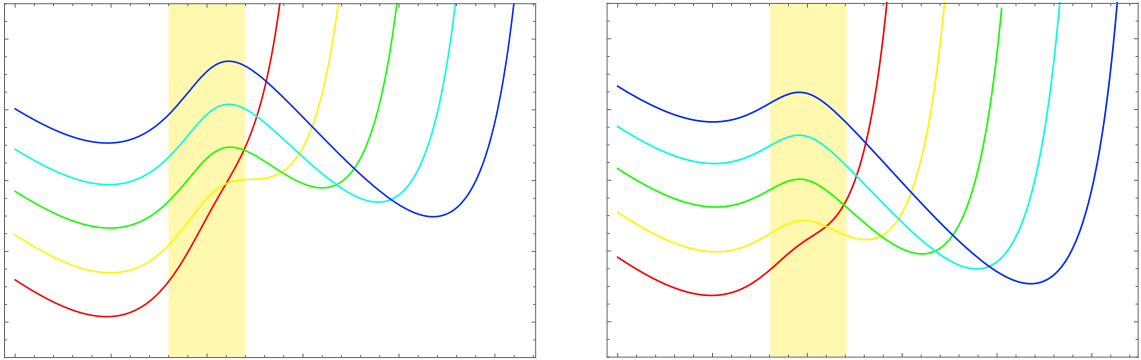
\begin{figure}
\begin{center}
\begin{tikzpicture}[scale=1]
\draw[yellow!40,fill=yellow!40] (2.4,0.26) rectangle (3.4,4.95);
\pgftext[at=\pgfpoint{0cm}{0cm},left,base]{\pgfuseimage{pic3}} 
\begin{scope}[xshift=8cm]
\draw[yellow!40,fill=yellow!40] (2.4,0.26) rectangle (3.4,4.95);
\pgftext[at=\pgfpoint{0cm}{0cm},left,base]{\pgfuseimage{pic4}} 
\end{scope}
\end{tikzpicture}
\caption{\footnotesize The off-shell generalized entropy for $R$ (left) and $R\cup\overline D$ for the operator quench model as a function of $v$  for several values of $t$ from red to blue. The region in yellow shows the region where the energy density is localized. For this plot, the parameters are: $k=0.1, c=10, h=5, h_D=1, \epsilon = 0.5, t_0=2/k$.}
\label{fig5}
\end{center}
\end{figure}

It is clear from the numerical analysis that for $R$ the maximum QES is at the far edge of $D$ at  $v=t_0+\epsilon$ whereas for $R\cup\overline D$ it lies inside the interval. This is simple to understand, the important terms in the generalized entropy that determine the $v$ coordinate of the QES when $v$ is in the $D$ interval are 
\EQ{
S_\text{gen}(v)\thicksim S_\text{BH}(v)+{\mathbb S}_D(v)\ ,
}
with ${\mathbb S}_D(v)$ defined for $R$ and $R\cup\overline D$ in \eqref{iuu}. Here, $S_\text{BH}(v)$ is a monotonically increasing function of $v$, due to the backreaction of $D$, whilst ${\mathbb S}_D(v)$ monotonically increases, for $R$, and decreases, for $R\cup\overline D$. Hence, for $R$ the maximum will lie at the edge of the $D$ interval at $v=t_0+\epsilon$ while for $R\cup\overline D$ there is a competition between the two terms and the position of the maximum in somewhere in the $D$ interval depending on the detailed form of the functions. These features are clear in the numerical analysis. 

In the case of $R$, the approximate coordinates of the QES at the recovery time are given by $v=t_0+\epsilon$ and \eqref{eq:U:QES_max_sol}.\footnote{One can arrive at the same result using \eqref{umm}.} The entropy of $R$ at the maximum is therefore given by
\EQ{
S_\text{max}(R)\approx S_\text{BH}(t_0+\epsilon)+S(D)+S_\text{rad}(R_2)=S_\text{BH}(t_0)+\Delta S_\text{BH}+S(D)+S_\text{rad}(R_2)\ .
\label{ucu}
}

We can now estimate the decoding complexity of $D$ from \eqref{hyd}, since  
\EQ{
S_\text{min}=S_{I_1}(R)=S_\text{BH}(t_0)+S_\text{rad}(R_2)\ ,
\label{eq:min_comp}
}
we have that
\EQ{
{\cal C}\thicksim\exp\big[(\Delta S_\text{BH}+S(D))/2\big]\ .
\label{eq:com_comp}
}
In the reversible case  $\Delta S_{\rm BH}= S_D$ so the decoding complexity for the diary is dominated by the exponential ${\cal C}\sim\exp(S(D))$, in agreement with the prediction in \cite{Brown:2019rox} and \cite{Yoshida:2017non}.
Notice that the irreversibility of the process of the black hole absorbing the diary increases the computational cost of decoding the diary.

\subsection{Island in the stream and python's lunch}

Let us now discuss a simplified way of deriving the complexity of decoding which involves the island in the stream formalism introduced in section \ref{s2}.
The procedure of shortening the bridge can be related to an off-shell sweep of the generalized entropy \cite{Brown:2019rox} $S_{I_\text{O-S}}(R)$ with an `off-shell' island $I_\text{O-S}$ whose end points are not necessary QES.
In the island in the stream formalism there is a natural way to define the generalized entropy off shell: simply use the islands in the stream formula \eqref{twm} but do not insist the  end points $\partial\tilde I$ lie in the subset $\partial R$. Varying one of the points $\partial\tilde I$ gives rise to a sweep.

Now consider the shockwave geometry case, and define a sweep corresponding to the generalized entropy with an off-shell island $\tilde{I}_\text{O-S}=[0^-, u_{O-S}]$.
\begin{center}
\begin{tikzpicture} [scale=0.8]
\filldraw[fill = Plum!10!white, draw = Plum!10!white, rounded corners = 0.2cm] (-3.2,1.3) rectangle (6.6,-3.3);
%
\draw[decorate,very thick,black!40,decoration={snake,amplitude=0.03cm}] (0,-2.5) -- (0,0.5);
\draw[dotted,thick] (0,0) -- (6,0);
\draw[dotted,thick,red] (0,-1) -- (6,-1);
\draw[dotted,thick,blue] (0,-2) -- (6,-2);
\draw[dotted,thick,OliveGreen] (3,-2.5) -- (3,0.5);
\draw[dotted,thick] (4,-2.5) -- (4,0.5);
\draw[very thick] (4,0) -- (0,0);
\filldraw[black] (0,0) circle (2pt);
\filldraw[black] (4,0) circle (2pt);
	\draw[very thick,red] (2,-1) -- (-0.2,-1);
	\filldraw[red] (2,-1) circle (4pt);
	\filldraw[red] (-0.2,-1) circle (4pt);
	\draw[red,->] (2,-1) -- (2,-0.7) -- (3.9,-0.7);
	\filldraw[OliveGreen] (3,-1) circle (2.5pt);
	\draw[OliveGreen] (3,-1) circle (4pt);
	\draw[very thick,blue] (2,-2) -- (4,-2);
	\filldraw[blue] (2,-2) circle (2pt);
	\filldraw[blue] (4,-2) circle (2pt);
	\node at (3,-2.8) {$t_0$};
		\node at (4,0.8) {$u$};
	\node[right] at (-3.2,0) {$R$};
	\node[right] at (-3.2,-1) {$\tilde{I}_\text{O-S}$};
	\node[right] at (-3.2,-2) {$R\ominus\tilde{I}_\text{O-S}$};
	\end{tikzpicture}
\end{center}

\begin{figure}[ht]
\begin{center}
\begin{tikzpicture} [scale=1] 
\draw[thick,black!40,->] (0,0) -- (0,5);
\draw[thick,black!40,->] (0,0) -- (7,0);
\draw[dotted] (4.05,1.75) -- (4.05,-0.2);
\draw[dotted] (5.5,1.2) -- (5.5,-0.2);
\draw[dotted] (-0.2,1.5) -- (6.5,1.5);
\draw[dotted] (4.1,2.25) -- (6.5,2.25);
\draw[thick,Blue] (0,2.2) -- (0.1,4.5) to[out=-50,in=170]  (4,1.5) -- (4.1,2.25) to[out=-45,in=170] (5.5,1.2) to[out=20,in=-145] (6.0,1.5);
\filldraw[Blue] (0,2.2) circle (2pt);
\filldraw[Blue] (0.1,4.53) circle (2pt);
\filldraw[Blue] (4,1.5) circle (2pt);
\filldraw[Blue] (4.1,2.25) circle (2pt);
\filldraw[Blue] (5.5,1.2) circle (2pt);
\node at (7.5,0) {$\tilde{u}_{O-S}$};
\node at (0,-0.4) {0};
\node at (4,-0.4) {$t_0$};
\node at (5.5,-0.4) {$u$};
\draw[<->] (6.2,2.25) -- (6.2,1.5);
\node at (7.7,1.8) {$S(D)+ \Delta S_\text{BH}$};
\node at (-0.8,2.2) {$S_\text{rad}(R)$};
\node at (-0.8,1.5) {$S_\text{min}$};
\node at (3.9,1.1) {$I_1$};
\node at (5.3,.85) {$I_2$};
\node at (0,5.3) {$S_{I_\text{O-S}}(R)$};
\end{tikzpicture}
\caption{\footnotesize The sweep $S_{I_\text{O-S}}(R)$ with $\tilde{I}_\text{O-S}=[0^-, u_{O-S}]$. The blue blobs are extrema of the generalized entropy. The sweep through the diary reveals a local maximum, a python's snack.}
\label{fig:sweep} 
\end{center}
\end{figure}
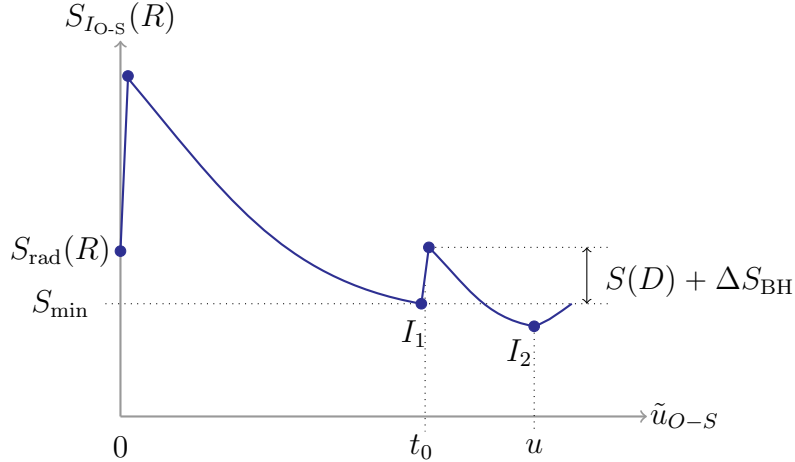

The sweep is shown in figure \ref{fig:sweep}, it  starts with an island just before the black hole is created, and then the entropy jumps by $S_\text{BH}(0)$ as $\tilde u$ crosses $0$. It then decreases until the diary is thrown in when $\tilde u=t_0$. 
Here, there is another minimum, which corresponds to the island saddle that gets stuck in front of the diary. Between this minimum and the dominating QES, there is a maximum which is given by the island saddle which sits just after the diary, which differs by the previous QES just by diary energy and entropy, as computed in \eqref{ucu} and \eqref{eq:min_comp}.

\section{Effect of the infalling system}
\label{sec:6}

The infalling system has some important consequences that sheds further light on some interesting issues. 

\subsection{Measurements outside affecting the inside}

It is a remarkable feature of the generalized entropy formalism that the island $I$ behind the horizon of a post Page-time black hole is in the entanglement wedge of $R$ and so the action of an operator acting in the effective theory on $I$ can be reconstructed at the microscopic level as an operator acting on the radiation $R$ \cite{
Penington:2019npb,Almheiri:2019qdq}. This seems to imply a breakdown of locality because it means that there are operators on the radiation that can manipulate the inside of the black hole. 

In order to shed more light on this apparent feature of the formalism, let us ask the following concrete question. Can any operation on the radiation $R=[0,u]$ affect an object $D$ that falls into the hole at some time $t_0>u$? It is important that $t_0>u$ because otherwise information allowing the recovery of $D$ itself will have been evaporated out and be encoded in the radiation and the question becomes moot. This answer is simple because we know that information cannot be recovered in the radiation until $D$ actually falls in; more precisely $I(R,\overline D)=0$ for $t_0>u$. Actually we have shown that $u$ can be slightly greater than $t_0$ to include the subset $R^\natural$ described in section \ref{s2}. So there is no correlation between the radiation $R$ and the purifier of $D$ meaning that the reduced state factorizes $\rho_{\overline DR}=\rho_R\otimes\rho_{\overline D}$.
This means that no operation on $R$, i.e.~any generalized measurement that can be represented as a quantum channel acting on $R$, can affect $D$. This is entirely consistent with the analysis in \cite{Yoshida:2019qqw}. This is consistent with the idea that after the Page time the island region inside of the black hole has been evaporated out into the radiation and is not in any meaningful sense behind the horizon any more.\footnote{More precisely, it is the information allowing the recovery of the Hawking partner modes that has been evaporated out of the black hole.} This breakdown of the semiclassical description might be signaled by a new kind of singularities, called `quantum singularities', discussed in \cite{Bousso:2022tdb}.

\subsection{Encoding of Hawking partners}

The entanglement monogamy problem and the implication for the existence of a firewall at the horizon are well known \cite{Almheiri:2012rt,Mathur:2009hf}. In short, the entropy of a Hawking mode $R_2$ is identified as due to its entanglement with its partner mode $\overline R_2$ behind the horizon in order to have the ingoing vacuum state around the horizon. Any disruption to this entanglement, for example, replacing the Unruh state by the Boulware state would lead to a divergence in the energy density at the horizon, i.e.~a firewall. On the other hand, for a black hole past the Page time, the Hawking mode must be correlated with the early radiation $R_1$. The Hawking mode cannot be correlated with both $\overline R_2$ and $R_1$ since this violates the monogamy of entanglement.

In the generalized entropy formalism, the monogamy problem evaporates. The purifier $\overline R_2$ is encoded in a redundant way in the early radiation $R_1$ and the remaining black hole. To be more specific, define $R_2=[t_0,t_1]$ (with $|t_1-t_0|\gg T^{-1}$) to be a small subset of Hawking modes emitted past the Page time. In this case, the modes in $R_2$ are correlated with the early radiation $R_1=[0,t_0]$ since the island $I_1=[0,t_0]$ dominates for $R_1$, giving
\EQ{
S(R_1)=S_{I_1}(R_1)=S_\text{BH}(t_0)\ ,
}
and the island $I_2=[0,t_1]$ for $R_1\cup R_2$, giving
\EQ{
S(R_1\cup R_2)=S_{I_2}(R_1\cup R_2)=S_\text{BH}(t_1)\ .
}
Hence,
\EQ{
I(R_1,R_2)=S_\text{rad}(R_2)+S_\text{BH}(t_0)-S_\text{BH}(t_1)=\frac{\xi+1}\xi S_\text{rad}(R_2)\ .
}
In the above, since $R_2$ is a small interval $S(R_2)=S_\varnothing(R_2)=S_\text{rad}(R_2)$. 

Note that the purifier $\overline R_2$ can also be located in the remaining black hole, or equivalently the late radiation $R_3=[t_1,\infty]$, and a subset of the early radiation $R_1$. The fact that some of the early radiation is needed follows from the fact $R_2$ has non-vanishing correlation with the early radiation $R_1$. 

If the partner modes $\overline R_2$ are encoded partially in the early radiation then this presents a potentially paradoxical situation for a freely falling observer who crosses the horizon. The observer who falls in at time $t_0$ needs to experience the Hawking modes $R_2$ and their partners $\overline R_2$ in the entangled Unruh state but if the partners are partially encoded in the early radiation how can that be unless there is some extreme form of non-locality?

However, as observed by Yoshida \cite{Yoshida:2018ybz,Yoshida:2019kyp,Yoshida:2019qqw,Yoshida:2021xyb}, it is important to take into account the backreaction created by $D$. In the present context, the infall of $D$ creates a competition between the islands $I_1$ and $I_2$ for $S(R)$ (as before $R=R_1\cup R_2$) and so now the correlation between the Hawking modes $R_2$ and the early radiation is
\EQ{
I(R_1,R_2)=\max\Big(0, \frac{\xi+1}\xi S_\text{rad}(R_2)-\Delta S_\text{BH}-S(D)\Big)\ ,
}
ignoring the log terms. So a small enough set of modes $R_2$ are now completed uncorrelated with the early radiation. The largest set of modes with this property is the set $R_2=R^\natural\cup R^\sharp$ defined earlier, precisely the minimum amount of radiation needed to recover $D$. This is clear from figure \ref{fig3}: when $I_1$ dominates for $R_1\cup R_2$ there are a set of partner modes behind the true horizon $\tilde U=0$ that do not lie in the island $I_1$. It is only when $R_2$ gets sufficiently large that the island $I_2$ dominates. So $D$ can experience the entangled state, the inertial vacuum, across the true horizon as it falls into the black hole and there is no need to invoke some extreme form of non-locality.

Given that $D$'s fate is to have its state scrambled and information evaporated out of the black hole, it is interesting to ask how long  does $D$ last inside the black hole from its own perspective rather than from the perspective of an observer outside the black hole who collects the radiation. In order to make this a meaningful question we first have to make $D$'s trajectory slightly time-like so we can talk about $D$'s proper time. A time-like geodesic in AdS$_2$ in KS coordinates has the form
\EQ{
\tilde U=\mu\frac{\sin(\tau/2)}{\sin(\tau_s-\tau/2)}\ ,\qquad \tilde{V}=\mu^{-1}\frac{\cos(\tau_s-\tau/2)}{\cos(\tau/2)}\ ,
}
where $\tau$ is the proper time and $\mu$ and $\tau_s$ are constants. Horizon crossing occurs at $\tau=0$ and the singularity is reached at $\tau=\tau_s$. For an almost null trajectory of $D$ (which has $\tilde V=1$), $\tau_s$ is small and the geodesic is approximately 
\EQ{
\tilde U\approx\frac{\tau}{2\tau_s-\tau}\ ,\qquad \tilde V\approx1\ .
}
valid in the region inside the black hole. 

Now that we can talk about $D$'s proper time, the question then is how to judge when $D$ has been scrambled and its information evaporated out of the black hole in its own frame. If we accept the premise that QFT modes that lie in the island (in this case the island $I_1$) have been partially evaporated out of the black hole and are now encoded in the radiation, since they lie in the entanglement wedge of the radiation $R_1$ emitted prior to $D$ falling in, and are no longer present inside the black hole then this puts a limit on the amount of the interior available to $D$. We propose that this sets a limit on how much of the interior $D$ can experience before being scrambled and evaporated out. The $\tilde U$ coordinate of the QES of the island $I_1$ is $\tilde U=4\lambda/3$ and $D$ will reach this at proper time $\tau=8\lambda\tau_s/3$. Hence, the ratio of proper times for an ultra-relativistic infalling object is
\EQ{
\frac{\text{time inside}}{\text{time to singularity}}\approx\frac{8\lambda}3\thicksim\frac{\Delta S_\text{BH}}{S_\text{BH}}\ ,
\label{eq:endu_time}
}
where the time to the singularity $\tau_s$ is defined in the classical spacetime.

\section{Discussion}
\label{s7}

We have considered how information is recovered in a model of an evaporating black hole in JT gravity where the backreaction problem for an infalling object is exactly solvable. In the case that we collect all the Hawking radiation from the beginning of evaporation, as discussed in section \ref{sec:HP_IITS}, we have seen how the generalized entropy formalism leads to information recovery in the way anticipated by Hayden and Preskill in their pioneering work \cite{Hayden:2007cs}. The formalism we developed allowed us to show that we can also recover the information from the radiation emitted after the diary falls in, as discussed in section \ref{sec:HP_late}. The fact that the information in the diary can be recovered in different subsets of the Hawking radiation can be seen as a consequence of the relationship between entanglement wedge reconstruction and quantum error correction.

In addition to previous work \cite{Hollowood:2021nlo}, in this paper we have also considered corrections to the entropy which are logarithmic in the black hole entropy and the shift in the black hole entropy, the latter arising from the backreaction of the diary. This refinement is necessary in order to evaluate the time needed for the information in the diary to start appearing in the radiation. This time was identified in \cite{Hayden:2007cs} as the scrambling time and, interestingly, it involves two terms \eqref{rhd2}: one term is proportional to $c \log S_\text{BH}/\Delta S_\text{BH}$, where $\Delta S_\text{BH}$ is the shift in the black hole entropy due to the backreaction of the infalling object, and can be interpreted as the usual scrambling time, whilst the other term is proportional to $\Delta S_\text{BH}-S(D)$, where $S(D)$ is the entropy of the diary. This other term has a thermodynamic origin; it is due to the fact that the process of the black hole absorbing the diary is an irreversible process if the first law $\Delta S_\text{BH} \ge S(D)$ is not saturated. Therefore, irreversibly delays the time at which the black hole returns the information in the diary to the radiation and so the black hole doesn't behave as a mirror anymore.

Irreversibly also increases the complexity of decoding the diary in the radiation. In section \ref{s5}, using the conjecture of \cite{Brown:2019rox}, which relates the complexity $\mathcal{C}$ to the size of the python's lunch, we found $\log \mathcal{C} \sim (\Delta S_\text{BH}+S(D))/2 \ge S(D)$, where $S(D)$ is the estimate based on reversible qubit models \cite{Yoshida:2017non}. 

In section \ref{sec:6} we considered the experience of an infalling observer as they pass through the horizon. Due to their backreaction, we found that the Hawking quanta emitted when the observer is crossing the horizon are not entangled with the early radiation, avoiding the entanglement monogamy paradox and granting a smooth experience for the observer through the horizon. However, we know that at a certain point the information of the infalling observer is accessible on the outside, which means that they have be scrambled in the Hawking radiation. This happens roughly when the observer crosses the island. We propose that this sets a limit on how much of the interior they can experience and calculated the endurance time for an ultra-relativistic infalling observer, which turns out to depend on the diary backreaction $\Delta S_\text{BH}$ as in equation \eqref{eq:endu_time}.

\vspace{0.5cm}
\begin{center}{\it Acknowledgments}\end{center}
\vspace{0.2cm}
TJH, AL and SPK acknowledge support from STFC grant ST/T000813/1. NT and ZG acknowledge the support of an STFC Studentship

\vspace{0.5cm}

\begin{center}
***
\end{center}

\vspace{0.5cm}

{\small For the purpose of open access, the authors have applied a Creative Commons Attribution (CC BY) licence to any Author Accepted Manuscript version arising.}

\appendix
\appendixpage

\section{Multipartite generalized entropy in the shockwave geometry}\label{app:A}

In this section, we show how the simple result for solution of the variational problem in the adiabatic limit \eqref{hys} follows in the general case when $R$ consists of an arbitrary set of intervals and there is an infalling object $D$ as described in section \ref{s2} in the adiabatic limit. In this section we assume that the matter theory consists of $c$ massless Dirac fermions, as the formula for the entropy for multiple intervals is known for this case \cite{Casini:2005rm}.

We will label the QES with $a,b,\dots$ while the intervals endpoints at $\mathscr I^+$ with $i,j,\dots$. We also use Greek letters to cover all the indices $\alpha=(a,i)$ ordered along a Cauchy slice. The expression for the generalized entropy simplifies in the adiabatic limit because, in this limit, the QES are close to the horizon. This is the would-be horizon $U=0$, for QES with $V_a<1$, and the true horizon $\tilde U=0$, for QES with $V_a>1$:\footnote{Remember that we have normalized our coordinates so that $D$ falls in along $V=\tilde V=1$.}
\EQ{
U_aV_a\ll1\ ,\quad\text{for}\quad V_a<1\ ,\qquad \tilde U_a\tilde V_a\ll1\ ,\quad\text{for}\quad V_a>1\ .
\label{get}
}
Our approach is to assume that this is true and then show that the solutions of the variational problem we find are consistent with this.

In the adiabatic approximation, and assuming that the QES are close to their appropriate horizons, the QES contribution to the generalized entropy simplifies to
\EQ{
S_\text{QES} = \sum_{V_a<1}S_\text{BH}(v_a)(1-2U_aV_a)+\sum_{V_a>1}S_\text{BH}(v_a)(1-2\tilde U_a\tilde V_a)\ .
}
The QFT contribution to the generalized entropy is 
\EQ{
S(\rho^\text{sc}_{R\cup I})= - \frac{c}{6} \sum_{\alpha < \beta} (-1)^{\alpha-\beta} \log (U_\alpha - U_\beta)-\frac{c}{12} \sum_\alpha \log U'_\alpha+ \frac{c}{12} \sum_a \log \frac{V'_aU'_a}{(1+U_a V_a)^2}\ .
\label{eq:S_QFT_w}
}
where $V'_a=V'(v_a)$ and $U'_\alpha=U'(u_\alpha)$ and for the QES we define the coordinate $u_a$ so that $U_a=-U(u_a)$. The above can also be expressed in terms of the $(\tilde U,\tilde V)$ coordinates in an identical way. Moreover, each term in the third sum is invariant under a M\"obius transformation by itself. 

We can now apply the assumption \eqref{get} and also the following expressions,
\begin{equation}
\frac{V'}{V}(t)=-\frac{U'}{U}(t)=2\pi T(t) \quad \text{for $t<t_0$} \, , \quad \frac{\tilde V'}{\tilde V}(t)=-\frac{\tilde U'}{\tilde U}(t)=2\pi T(t) \quad \text{for $t>t_0$} \, ,\\
\end{equation}
valid in the adiabatic limit, for writing:
\EQ{
S(\rho^\text{sc}_{R\cup I})&= - \frac{c}{6} \sum_{\alpha < \beta} (-1)^{\alpha-\beta} \log (U_\alpha - U_\beta)-\frac\cc{12}\sum_{|U_\alpha|>1}\log |U_\alpha|\\[5pt] &-\frac\cc{12}\sum_{|U_\alpha|<1} \log|\tilde U_\alpha|
+ \frac{c}{12} \sum_{V_a<1} \log V_aU_a+\frac c{12}\sum_{V_a>1}\log \tilde V_a\tilde U_a\ ,
\label{eq:S_QFT_w} 
}
up to some subleading terms of order $\cc\log T$. In the above, we have used the approximation $U=\tilde U-\lambda$ so that $U'\approx \tilde U'$ valid when $|U|<1$. 

Varying the generalized entropy with respect to $V_a$ for QES with $V_a<1$ and $\tilde V_a$ for QES with $V_a>1$, yields the approximate solutions
\EQ{
V_a U_a = \frac{c}{48 S_\text{BH}(v_a)} \ll 1\ ,\qquad \tilde V_a\tilde U_a = \frac{c}{48 S_\text{BH}(v_a)} \ll 1\ ,
\label{yes}
}
respectively. Since $S_\text{BH}\gg\cc$ in the adiabatic approximation, this justifies our earlier assumption \eqref{get}. These conditions have the approximate solution
\EQ{
v_a=u_a-\frac1{2\pi T(u_a)}\log\frac{S_\text{BH}(u_a)}\cc\ ,
}
where the second term is assumed to be small compared with the size of the intervals which means for any interval $R_j\subset R$, we have $S_\text{rad}(R_j)\gg \cc\log S_\text{BH}/\cc$.

Now let us turn to extremization with respect to $U_a$. We split the case when the QES is in front of $D$, i.e.~$v_a<t_0$, into two cases, depending whether $u_a \lessgtr t_0$. Taking the case $u_a < t_0$ first. Differentiating the generalized entropy with respect to $U_a$ and using \eqref{yes} we have
\begin{equation}
{v_a < t_0\atop u_a \lesssim t_0}: \qquad \frac14 +\sum_{\alpha \neq a} (-1)^{a-\alpha} \frac{U_a}{U_a - U_\alpha}=0 \ ,
\label{kek}
\end{equation}
which is familiar form the analysis in \cite{Hollowood:2021nlo}. Solutions to this equation are simple because there is a hierarchy of scales amongst the coordinates because the intervals in $R$ are large compared with the thermal scale, so $u_i>u_j$ implies $|U_i|\ll |U_j|$ \footnote{Notice that, since the diary is small $\lambda \ll 1$, we have that the hierarchy between the KS coordinates will hold true also when $u_a \sim t_0$, as indicated in equation \eqref{kek}, i.e. until $|\tilde{U}_a| \gg \lambda$.}. The implication is that each QES coordinate lies close to an end point in the sense that $U_a = \kappa |U_j|$ for a numerical factor $\kappa = O(1)$ and the sum is well approximated by
\begin{equation}
{v_a < t_0\atop u_a \lesssim t_0}: \qquad \frac14 - \frac{U_a}{U_a - U_j}=0 \quad \implies \quad U_a = \frac13 |U_j| \ .\label{wee}
\end{equation}
There is a subtlety here. If we order the coordinates $\{U_\alpha\}=\{U_j\}\cup\{U_a\}$ in order of magnitude $|U_\alpha|$ then the number of coordinates that have $|U_\alpha|\ll |U_j|$ must be even. The case when the number is odd also give solutions but these always turn out to be maxima of the generalized entropy and we are not interested in these.

Now consider the case with $v_a<t_0$ but now with $u_a > t_0$. We have the same equation \eqref{kek} but the presence of $D$ complicates the discussion since we have that all the radiation endpoints with $u_i > t_0$ have coordinates $U_i \sim - \lambda$ and we expect $U_a = O(\lambda)$, indeed we will consider an ansatz of the form $U_a = \kappa \lambda$.  In this case we can neglect in  \eqref{kek} all the terms with $|U_i| \gg \lambda$ while all the other ones have an identical contribution. However, considering again the case where we have an even number of coordinates $\lambda \sim |U_\alpha|\ll 1$, we get that all these terms will exactly cancel each other leaving us with:
\begin{equation}
{{v_a< t_D}\atop{\hat{u}_a > t_D}}: \qquad \frac14 - \frac{U_a}{U_a + \lambda}=0 \quad \implies  \quad U_a = \frac13 \lambda \ .
\end{equation}

Finally, consider the case where the QES is after $D$, $v_a>t_0$.
The extremization with respect to $U_a$ leads to:
\begin{equation}
v_a> t_0: \qquad \frac14 +\sum_{\alpha \neq a} (-1)^{a-\alpha} \frac{\tilde{U}_a}{U_a - U_\alpha}=0 \, . \label{eq:extr_after_1}
\end{equation}
Again, the presence of $D$ affects the discussion, but the key point is that it does not change the hierarchy amongst the coordinates, in the sense that, for $u_i>u_j$,
\EQ{
	U_i-U_j \approx \begin{cases} -U_j &u_j<t_0\ ,\\ -\tilde U_j & u_j>t_0\ , \end{cases}
}
where we have used $U_j = -\lambda+\tilde{U}_j$ for $u_j>t_0$.
So, the hierarchy of scales amongst the $U_\alpha$ coordinates can still be exploited to give a solution
\EQ{
v_a> t_0: \qquad  \tilde U_a=\frac13|\tilde U_j|\ .
}

It is noteworthy that when $u_j > t_D$ there are two saddles that have $v_a\gtrless t_0$. When the QES is in front of $D$, we have $U_a = \lambda/3$ and, approximately,
\EQ{
v_a = t_0 - \frac{1}{2 \pi T(t_0)}\log  \frac{\Delta S_D}{24 c} \ .
\label{pop1}
}
The other saddle has
\EQ{
v_a = u_j - \frac1{2\pi T(u_j)}\log\frac{S_\text{BH}(u_j)}c > t_0\ .
\label{pop2}
}

Using \eqref{pop1} and \eqref{pop2}, we can compute the logarithmic corrections to entropy of the saddle. Each QES contributes a correction which gives \eqref{twm2} plus a correction specific for the case where the QES gets stuck before the diary ($\delta_{u_{\partial\tilde{I}} , t_0}=1$)
\begin{equation}
S_I = \sum_{\partial \tilde{I}}S_\text{BH}(u_{\partial \tilde{I}})+ S_\text{rad}(R \ominus \tilde{I}) - \frac{c}{24} \sum_{\partial \tilde{I}} \left[ \log \frac{S_{\text{BH}}(u_{\partial \tilde{I}})}{c} -3 \delta_{u_{\partial\tilde{I}} , t_0} \, \log \frac{\Delta S_\text{BH}}{S_\text{BH}(t_0)} \right] .
\end{equation}
This is a refinement of \eqref{twm}.

\end{document}